\documentclass[12pt]{article}
\usepackage{mydraft}
\usepackage[ruled,vlined]{algorithm2e}

\def \varthetab {\boldsymbol{\vartheta}}
\def \Bcalb {\boldsymbol{\mathcal{B}}}
\newcommand{\zeronorm}[1]{\lVert#1\rVert_0}
\begin{document}

\newcommand{\blind}{0}

\if0\blind{
\title{\bf Matrix-Response Generalized Linear Mixed Model with Applications to Longitudinal Brain Images}
\author{
Zhentao Yu, Jiaqi Ding, Guorong Wu, and Quefeng Li\\[4pt]
\textit{Department of Biostatistics, Gillings School of Global Public Health,}\\
\textit{University of North Carolina at Chapel Hill}\\[2pt]
\textit{Department of Computer Science, College of Arts and Sciences,}\\
\textit{University of North Carolina at Chapel Hill}\\[4pt]
}
\maketitle
} \fi



\begin{abstract} {Longitudinal brain imaging data facilitate the monitoring of structural and functional alterations in
    individual brains across time, offering essential understanding of dynamic neurobiological mechanisms. Such data
    improve sensitivity for detecting early biomarkers of disease progression and enhance the evaluation of intervention
    effects. While recent matrix-response regression models can relate static brain networks to external predictors,
    there remain few statistical methods for longitudinal brain networks, especially those derived from high-dimensional
    imaging data. We introduce a matrix-response generalized linear mixed model that accommodates longitudinal brain
    networks and identifies edges whose connectivity is influenced by external predictors. An efficient Monte Carlo
    Expectation–Maximization algorithm is developed for parameter estimation. Extensive simulations demonstrate
    effective identification of covariate-related network components and accurate parameter estimation. We further
    demonstrate the usage of the proposed method through applications to diffusion tensor imaging (DTI) and functional
    MRI (fMRI) datasets.}

\end{abstract}
\noindent%
\vfill
\section{Introduction}
The Human Connectome Project (HCP) is a large-scale international effort to map the structural and functional
connectivity networks of the human brain. Network models are widely used in brain connectivity research
\citep{bassett2018models}. Such studies acquire multimodal neuroimaging data—including Magnetic Resonance Imaging (MRI),
Positron Emission Tomography (PET), and Diffusion Tensor Imaging (DTI)—from multiple subjects to characterize the
brain’s functional and structural organization \citep{babaeeghazvini2021review}. For each subject, these data can be
used to construct a connectivity network, with nodes representing regions of interest (ROIs) and edges encoding
functional or structural relationships between regions. Alterations in these networks are associated with
neurodegenerative disorders, including Alzheimer’s disease, Lewy body dementia, and Parkinson’s disease. A central
challenge in neuroscience is to identify biomarkers that capture or predict network changes, enabling earlier diagnosis
and intervention. The recent emergence of large imaging data consortia, such as ADNI (https://adni.loni.usc.edu) and UK
Biobank (https://www.ukbiobank.ac.uk/), has given researchers access to large-scale longitudinal brain-imaging datasets,
which are preferable to traditional cross-sectional snapshots acquired at a single time point. These longitudinal
datasets are invaluable for detecting age-related cognitive decline, tracking the progression of neurodegenerative
diseases, mapping the development of brain function in youth, and addressing numerous other key questions in
neuroscience.

Generalized linear mixed models (GLMMs) are widely used for longitudinal data. In GLMMs, fixed
effects capture population-level trends, whereas random effects account for between-subject variability. However,
standard GLMMs do not directly accommodate matrix-valued responses and thus cannot be used to relate brain networks to
external predictors. Recent advances in matrix-response regression address this limitation. \cite{zhang2023connectivity}
proposed matrix-response generalized linear models for cross-sectional brain images, extending traditional generalized
linear models to matrix-valued responses and showing that, by exploiting low-rank and sparse structures in the
coefficient matrices, their estimators are consistent; they also demonstrated superior empirical performance over
methods that ignore matrix structure. \cite{zhou2022network} developed graph Laplacian regression models that perform
local smoothing on graph Laplacians. While applicable to brain networks, their methods target a specialized class (graph
Laplacians) rather than general network matrices. Moreover, neither of the above methods handles longitudinal imaging
data. \cite{zhao2024longitudinal} proposed a longitudinal regression model for covariance-matrix outcomes. Although it
accommodates longitudinal imaging data, the response is a scalar derived from eigenpairs of covariance matrices rather
than a general matrix. Consequently, it cannot directly identify features that drive changes in whole-brain network
structure. More details of these existing methods can be found in Section \ref{sec:exist-matr-resp}.

In this article, we introduce a class of matrix-response generalized linear mixed models (MR-GLMMs) for longitudinal
brain networks, where responses are time-indexed square matrices. We assume that after transformation of canonical link
functions, their conditional means can be written as a linear combination of time-varying covariates with matrix-valued
coefficients plus a matrix-valued intercept that accommodates random effects. We assume that the matrix-intercept term
is low-rank and the matrix-coefficients are sparse. Then, we propose to solve a constrained likelihood maximization
problem to obtain estimators of parameters of MR-GLMMs. The main challenge is that the likelihood function involves
evaluating high-dimensional integrals, which are computationally challenging. To address this, we develop an efficient
Monte Carlo Expectation–Maximization (MCEM) algorithm to compute the estimators. In the E-step, we employ
Metropolis-within-Gibbs (MwG) schemes previously developed in \cite{heiling2024glmmPen} to approximate the required
integrals. We conduct extensive simulations to evaluate variable selection and estimation accuracy of our method and
compare it against element-wise penalized GLMMs. Across scenarios, our approach consistently outperforms these
alternatives. We further analyze two longitudinal neuroimaging datasets, illustrating the model’s ability to identify
key biomarkers.

The rest of the paper is organized as follows. In Section \ref{sec:matr-resp-gener}, we introduce the proposed MR-GLMM
and compare it with other matrix-response regression models in the literature. In Section \ref{sec:monte-carlo-expect},
we amplify the details of the Monte Carlo Expectation-Maximization algorithm for solving parameters in our proposed
model and tuning parameter selections. Our method's effectiveness is demonstrated in Section \ref{sec:simulation}
through extensive simulation studies comparing with the element-wise penalized GLMM approaches across various
scenarios. In Section \ref{sec:appl-brain-imag}, we apply our method to two real-world neuroimaging datasets: a DTI
dataset from ADNI examining age, sex, APOE4, and diagnostic status effects on structural connectivity, and a fMRI
dataset from the HCP investigating task-induced changes in resting-state networks. 

\section{Matrix Response Generalized Linear Mixed Model}
\label{sec:matr-resp-gener}
\subsection{Notation}
For a matrix $\boldsymbol{A} \in \mathbb{R}^{d_1\times d_2}$, let $\boldsymbol{A}_{ij}$ denote the $(i,j)$-th element,
and let $\boldsymbol{A}_{i.}$ and $\boldsymbol{A}_{.j}$ represent its $i$-th row and $j$-th column, respectively. Let
$\|\boldsymbol{A}\|_2 = \lambda_{\max}(\sqrt{\boldsymbol{A}^{\top}\boldsymbol{A}})$ and
$\|\boldsymbol{A}\|_F=\sqrt{\sum_{i,j}|\boldsymbol{A}_{ij}|^2}$ denote the $L_2$ and Frobenius norms of
$\boldsymbol{A}$. Let $\text{SVD}_r(\boldsymbol{A}) = [\boldsymbol{U}, \boldsymbol{\Sigma}, \boldsymbol{V}]$ denote the
rank-$r$ truncated Singular Value Decomposition (SVD) of matrix $\boldsymbol{A}$ such that
$\boldsymbol{A} = \boldsymbol{U} \boldsymbol{\Sigma} \boldsymbol{V}^{\top}$, where
$\boldsymbol{\Sigma} \in \mathbb{R}^{r \times r}$ is a diagonal matrix containing the $r$ largest singular values of
$\boldsymbol{A}$, and $\boldsymbol{U} \in \mathbb{R}^{d_1 \times r}$, $\boldsymbol{V} \in \mathbb{R}^{d_2 \times r}$
contain the corresponding left and right singular vectors, respectively. For a tensor
$\mathcal{B} \in \mathbb{R}^{d_{1} \times d_{2} \times d_{3}}$, let $\mathcal{B}_{i,j,k}$ denote its $(i,j,k)$-th entry,
$\mathcal{B}_{i,j,:}$ denote the $(i,j)$-th tube fiber, and $\mathcal{B}_{:,:,k}$ denote the $k$-th frontal slice of
$\mathcal{B}$. Let $\|\mathcal{B}\|_F = \sqrt{\sum_{ijk}\mathcal{B}_{ijk}^2}$ denote its Frobenious norm and
$\|\mathcal{B}\|_0$ denote the number of nonzero elements in $\mathcal{B}$.

\subsection{Model Settings}
For a sample of $N$ individuals, suppose we observe independent longitudinal network data represented by adjacency
matrices $\boldsymbol{A}_{it}\in \mathbb{R}^{n\times n}$, where $1\leq i\leq N$ indexes the subjects and $1\leq t\leq T_i$ denotes the time points, with $T_i$ being the number of observations for subject $i$. Each $n \times n$ adjacency matrix represents a network with $n$ nodes. Additionally, we observe time-varying covariates $\boldsymbol{x}_{it} = (x_{it1},...,x_{itp})^{\top}\in \mathbb{R}^p$ for each individual. We assume that, conditional on $\boldsymbol{x}_{it}$, the adjacency matrix $\boldsymbol{A}_{it}$ follows a multiplicative exponential family distribution with a canonical link function, whose density function has the form
\begin{equation*}
  f(\boldsymbol{A}_{it}|\boldsymbol{x}_{it}) = \prod_{j\neq j'}^n h(A_{it,jj'}) \exp\left( A_{it,jj'} \eta_{it,jj'} - \psi(\eta_{it,jj'}) \right),
\end{equation*}
where $\eta_{it,jj'}$ is the $(j,j')$-th element of $\boldsymbol{\eta}_{it}=g(\boldsymbol{\mu}_{it})$, $\boldsymbol{\mu}_{it}=\mathbb{E}(\boldsymbol{A}_{it}|\boldsymbol{x}_{it})$, $g(\cdot)$ is a commonly-used link function in GLM applied element-wise, and $\psi(\cdot)$ is the cumulant function satisfying $\psi'(\cdot)=g^{-1}(\cdot)$. We assume that the link function takes the form
\begin{equation}
  \label{eq:link}
  g(\boldsymbol{\mu}_{it}) = \boldsymbol{\Theta} + \boldsymbol{\theta}_i +  \mathcal{B}\times_3 \boldsymbol{x}_{it}, 
\end{equation}
where $\boldsymbol{\Theta} \in \mathbb{R}^{n\times n}$ is the fixed intercept matrix that characterizes the
population-level connectivity, $\boldsymbol{\theta}_i \in \mathbb{R}^{n\times n}$ is the random intercept matrix for
subject $i$, and each element $\theta_{i,jk}$ is assumed to follow $\mathcal{N}(0, \sigma_{\theta, jk}^2)$. We define
$\boldsymbol{\Sigma}_{\theta} \in \mathbb{R}^{n\times n}$ as the matrix of variances where
$[\boldsymbol{\Sigma}_{\theta}]_{jk} = \sigma_{\theta, jk}^2$, and $\mathcal{B} \in R^{n\times n \times p}$ as the slope
tensor that defines the effects of covariates on the connectivity matrix, and
${\mathcal{B}\times_3}\x_{it}=\sum_{l=1}^p x_{itl}\mathcal{B}_{:,:,l}$, where ${x_{itl}}$ is the $l$-th element of
$\x_{it}$. We assume that the population-level connectivity matrix $\boldsymbol{\Theta}$ has a low-rank structure, a
common assumption in brain network modeling \citep{zhang2023connectivity}. Under this assumption, we perform the
Burer-Monteiro factorization of $\boldsymbol{\Theta}$ by writing it as
$\boldsymbol{\Theta} = \boldsymbol{U}\boldsymbol{V}^{\top}$ \citep{burer2003nonlinear}, where
$\boldsymbol{U}, \boldsymbol{V} \in \mathbb{R}^{n \times r}$ and $r$ denotes the rank of $\boldsymbol{\Theta}$. This
reparameterization eliminates the need for computationally expensive SVD calculations at each iteration and represents a
standard approach in optimization literature for enforcing low-rank constraints. Furthermore, we assume that
$\mathcal{B}$ is exactly sparse in the sense that many of its elements are exactly zero. This sparsity assumption posits
that covariate effects are confined to specific subsets of connections, consistent with findings from empirical
neuroscience research. Then, the parameters of interests are
$\boldsymbol{\vartheta} = (\boldsymbol{U},\boldsymbol{V},\mathcal{B})$. 
We denote $\ell(\boldsymbol{\theta})$ as the observed marginal full log-likelihood of the form
\begin{equation}
  \label{eq:obs-likelihood}
  l(\varthetab)=\sum_{i=1}^N \log\int f(\boldsymbol{A}_i| \boldsymbol{x}_i,\boldsymbol{\theta}_i;\varthetab)\phi(\boldsymbol{\theta}_i)d\boldsymbol{\theta}_i,
\end{equation}
where $f(\boldsymbol{A}_i| \boldsymbol{x}_i,\boldsymbol{\theta}_i;\varthetab)=\prod_{t=1}^{T_i}f(\boldsymbol{A}_{it}| \boldsymbol{x}_{it},\boldsymbol{\theta}_i;\varthetab)$, and $\phi(\thetab_i)$ denotes
the normal density of $\thetab_i$. To estimate $\varthetab$, we
propose to solve the following optimization problem:
\begin{equation}
  \label{eq:opt}
  \min_{\varthetab} ~ -l(\varthetab)+ \gamma\fnorm{\boldsymbol{U}^{\top}\boldsymbol{U}-\boldsymbol{V}^{\top}\boldsymbol{V}}^2, \text{ subject to } \zeronorm{\mathcal{B}} \le sn^2,
\end{equation}
where $\gamma$ and $s$ are two positive tuning parameters. In (\ref{eq:opt}), we incorporate a regularization term
$\fnorm{\boldsymbol{U}^{\top}\boldsymbol{U}-\boldsymbol{V}^{\top}\boldsymbol{V}}^2$ to enforce uniqueness in the factorization $\boldsymbol{\Theta}=\boldsymbol{U}\boldsymbol{V}^{\top}$, a technique also employed in other neuroimaging approaches \citep{zhang2023connectivity}. Moreover, we add the constraint
$||\mathcal{B}||_0 \leq sn^2$ to ensure rendering a sparse estimator of $\mathcal{B}$. The $L_0$-constraint is known to introduce less bias into the estimator compared to other constraints, such as the $L_1$-constraint.

\subsection{Existing Matrix-response Regression Models}
\label{sec:exist-matr-resp}
\cite{zhang2023connectivity} proposed cross-sectional matrix-response generalized linear models for studying brain
networks. Their model assumes that $\{\boldsymbol{A}_i\}_{i=1}^N$ are i.i.d. networks represented by $n\times n$
matrices, where $\boldsymbol{A}_i\in \mathbb{R}^{n\times n}$. These networks are regressed on a set of covariates using
generalized linear models. Specifically, let $\boldsymbol{A}_{i,jj'}$ denote the $(j,j')$-th element of
$\boldsymbol{A}_i$, and $\x_i\in \mathbb{R}^p$ be the covariate vector. They assume that the conditional density of
$\boldsymbol{A}_i|\x_i$ follows the canonical form of the exponential family:
\begin{equation*}
  f(\boldsymbol{A}_{i}|\boldsymbol{x}_{i}) = \prod_{j\neq j'}^n h(\boldsymbol{A}_{i,jj'}) \exp\left( \boldsymbol{A}_{i,jj'} \eta_{i,jj'} - \psi(\eta_{i,jj'}) \right),
\end{equation*}
where $\eta_{jj'}$ is the $(j,j')$-th element of
$\boldsymbol{\eta}_i=g(\boldsymbol{\mu}_{i}) = \boldsymbol{\Theta} + \sum_{l=1}^{p} x_{il}\mathcal{B}_{:,:,l}$ that
$\mub_i=\E(\boldsymbol{A}_i|\x_i)$, $x_{il}$ is the $l$-th element of $\x_i$ and $\mathcal{B}_{:,:,l}$ is the $l$-th frontal slice
of $\mathcal{B}$, $g(\cdot)$ is the link function, and $\psi'(\cdot)=g(\cdot)^{-1}$. To deal with high-dimensional
networks, they assumed that $\Thetab$ is low-rank and $\mathcal{B}$ is sparse. Then, they proposed to solve a penalized
GLM problem by imposing a low-rank decomposition on $\Thetab$ and sparsity-induced penalty on $\mathcal{B}$. They showed
that the resulting estimators have desirable statistical properties and satisfactory numerical performance through
extensive numerical studies. Their method is applicable to study cross-sectional brain networks. However, their method
is not applicable for longitudinal brain networks, which are the main objects of interest of our proposed method. 

\cite{zhou2022network} considered cross-sectional graph Laplacian regression models. For a graph with $n$ nodes and
bounded non-negative edge weights $w_{ij}$, its graph Laplacian is defined as $\A=(a_{ij})\in \mathbb{R}^{n\times n}$,
where $a_{ij}=-w_{ij}$ for $i\neq j$ and $a_{ii}=\sum_{k\neq i} w_{ik}$. Define the space of graph Laplacians as 
\(\mathcal{A}_n = \{\boldsymbol{A} \in \mathbb{R}^{n \times n} : \boldsymbol{A} = \boldsymbol{A}^\top,\, \boldsymbol{A}\boldsymbol{1}_n = \boldsymbol{0}_n\}\), equipped with a metric
\begin{equation*}
d(\boldsymbol{A}_1,\boldsymbol{A}_2)\;=\;\|\boldsymbol{A}_1-\boldsymbol{A}_2\|_F \;=\;\{\mathrm{tr}[(\boldsymbol{A}_1-\boldsymbol{A}_2)^\top(\boldsymbol{A}_1-\boldsymbol{A}_2)]\}^{1/2}.
\end{equation*}
Let $\X\in \mathbb{R}^p$ be the random covariate vector and $\x\in \mathbb{R}^p$ be a deterministic vector. For a graph
Laplacian $\A$, their model aims to find the graph Laplacian that minimizes the conditional Fréchet mean, i.e.,
\[
  m(\boldsymbol{x})\;=\;\arg\min_{\boldsymbol{\omega}\in\mathcal{A}_n}\;
  \mathbb{E}[d^{2}(\boldsymbol{A},\boldsymbol{\omega})\mid \boldsymbol{X}=\boldsymbol{x}]. 
\]
They proposed approximating such a conditional mean function using a function that minimizes the weighted squared
distance, i.e.,
\[
  m_G(\boldsymbol{x}) = \arg\min_{\boldsymbol{\omega} \in \mathcal{A}_n}
  \mathbb{E}[s_G(\boldsymbol{x})\,d^{2}(\boldsymbol{A},\boldsymbol{\omega})],
\]
where
$s_G(\boldsymbol{x}) = 1 + (\boldsymbol{X} - \boldsymbol{\mu})^{\top} \boldsymbol{\Sigma}^{-1}
(\boldsymbol{x} - \boldsymbol{\mu})$, $\boldsymbol{\mu} = \mathbb{E}(\boldsymbol{X})$, and
$\boldsymbol{\Sigma} = \operatorname{Var}(\boldsymbol{X})$. In practice, with i.i.d graph Laplacians $\{\A_i \}_{i=1}^n$,
an estimator of $m_G(\x)$ can be obtained by solving
\[
  \hat{m}_G(\boldsymbol{x}) = \arg\min_{\boldsymbol{\omega} \in \mathcal{A}_n} \frac{1}{N} \sum_{i=1}^{N}
  s_G^{(i)}(\boldsymbol{x})\,d^{2}(\boldsymbol{A}_i,\boldsymbol{\omega}),
\]
where $s_G^{(i)}(\boldsymbol{x})=1 + (\boldsymbol{X}_i - \bar{\boldsymbol{X}})^{\top} \hat{\boldsymbol{\Sigma}}^{-1}
(\boldsymbol{x} - \bar{\boldsymbol{X}})$, $\X_i$ is the covariate vector of the $i$-th subject, $\bar{\X}=N^{-1}
\sum_{i=1}^{N} X_i$, and $\hat{\Sigmab}$ is the sample covariance of $\Sigmab$. When $m(\x)$ is a smooth function, they
further proposed a locally smoothed estimator by solving
\begin{equation*}
    \hat{m}_{L}(\boldsymbol{x}) = \arg\min_{\boldsymbol{\omega} \in \mathcal{A}_n} \frac{1}{N} \sum_{i=1}^{N}
  s_{L}^{(i)}(\boldsymbol{x},h)\,d^{2}(\boldsymbol{A}_i,\boldsymbol{\omega}),
\end{equation*}
where
$\hat{s}_L^{(i)}(\boldsymbol{x},h) = \hat{\delta}^{-1}[K_h(\boldsymbol{X}_k - \boldsymbol{x})\, \{1 - (\boldsymbol{X}_k
- \boldsymbol{x})^\top \hat{\boldsymbol{\Xi}}^{-1} \hat{\boldsymbol{\mu}}_1 \}]$, $K_h(u)=h^{-1}K(u/h)$ is a kernel
function with bandwidth $h$,
$\hat{\delta}=\hat{\mu}_0 - \hat{\boldsymbol{\mu}}_1^\top \hat{\boldsymbol{\Xi}}^{-1}
\hat{\boldsymbol{\mu}}_1$, and $\hat{\mu}_{0}$, $\hat{\mub}_1$ and $\hat{\mub}_2$ are moment estimators of ${\mu}_0 = \mathbb{E}[K_h(\boldsymbol{X} - \boldsymbol{x})]$,  
$\boldsymbol{\mu}_1 = \mathbb{E}[K_h(\boldsymbol{X} - \boldsymbol{x})(\boldsymbol{X} - \boldsymbol{x})]$,  
$\boldsymbol{\Xi} = \mathbb{E}[K_h(\boldsymbol{X} - \boldsymbol{x})(\boldsymbol{X} - \boldsymbol{x})(\boldsymbol{X} -
\boldsymbol{x})^\top]$, respectively. 

Compared with our method, the approach of \cite{zhang2023connectivity} targets a more specialized class of networks, namely graph Laplacians, whereas our matrix-response model imposes fewer restrictions on network structure. Beyond this scope difference, there are several notable distinctions. First, their method remains a cross-sectional regression model and thus cannot be applied to longitudinal imaging data; it also makes it difficult to accommodate between-subject heterogeneity. Second, their local smoothing approach does not extend well to high-dimensional imaging data, as kernel smoothing is known to suffer from the curse of dimensionality. Finally, although the estimated conditional mean function enables prediction of graph Laplacians, it is less interpretable than our parametric additive model.

\cite{zhao2024longitudinal} developed a longitudinal regression model for analyzing covariance matrix outcomes. The
model identifies principal components of covariance matrices that are associated with the covariates, estimate
regression coefficients, and discovers within-subject variation in covariance matrices. In particular, they considered
longitudinal covariance matrices $\A_{it}$ for $1\leq i\leq N$ and $1\leq t\leq T_i$ and proposed the following model:
\[
  \log(\boldsymbol{\gamma}^{\top} \boldsymbol{A}_{it} \boldsymbol{\gamma}) = \beta_{0} +
  u_{i}+\boldsymbol{x}_{1i}^{\top} \boldsymbol{\beta}_{1} + \boldsymbol{x}_{2it}^{\top} (\boldsymbol{\beta}_{2} +
  \boldsymbol{\nu}_{i}),
\]
where $\gammab\in \mathbb{R}^p$ is the principal component vector, $\beta_0$ is the fixed intercept, $u_i$ is the random
intercept, $\x_{1i}$ is the time-invariant covariate with a fixed-effect coefficient $\betab_1$, $\x_{2it}$ is the
time-varying covariate with random-effect coefficient $\betab_2+\nub_i$. This model can be regarded as a generalized
linear mixed model with a logarithm link function applied to the univariate response
$\boldsymbol{\gamma}^{\top} \boldsymbol{A}_{it} \boldsymbol{\gamma}$. It aims to find the association between covariates
and the covariance matrices. They proposed to obtain estimators of those parameters by maximizing the
hierarchical-likelihood function with identifiability constraint. They proved that the resulting estimators are
asymptotically consistent and the corresponding covariance matrix estimator is efficient in both low and
high-dimensional settings.

Although their model can handle longitudinal brain images, it differs from ours in several fundamental ways. First, in
our model (\ref{eq:link}), the entire matrix serves as the response, and the coefficient tensor $\mathcal{B}$ fully
encodes how the covariates influence each matrix entry (i.e., each network edge). By contrast, their model's response is
an eigenvalue-based metric rather than the matrix itself. Consequently, the two models are intrinsically
different. Moreover, the dimensionality of our model's parameter space is substantially higher than theirs, posing much
greater computational challenges for developing efficient algorithms to solve our problem. In the next section, we give
more details of our algorithm.

\section{Monte Carlo Expectation Maximization Algorithm}
\label{sec:monte-carlo-expect}
We propose a Monte Carlo Expectation–Maximization (MCEM) algorithm to solve the optimization problem in
(\ref{eq:opt}). The observed likelihood function in (\ref{eq:obs-likelihood}) requires evaluating an expectation with
respect to the distribution of $\thetab_i$, which is essentially a high-dimensional integration problem and
computationally challenging.  In the E-step of the algorithm, we propose to approximate the expectation using Monte
Carlo sampling approach and minimize the resulting Q-function \citep{gilks1995adaptive} in the subsequent M-step. Then,
we iterate between the E-step and M-step.

At the $h$-th iteration of the MCEM algorithm, the complete log-likelihood function can be written as
\begin{equation}
  \begin{aligned}      \log(L_c(\d_{c};\varthetab|\boldsymbol{d}_o,\varthetab^{(h)}))
    &=\log\prod_{i=1}^N f(\boldsymbol{A}_i,\boldsymbol{x}_i,\thetab_i|\boldsymbol{d}_o,\varthetab^{(h)})\phi(\boldsymbol{\theta}_i)\\   &=\sum_{i=1}^N\log f(\boldsymbol{A}_i,\boldsymbol{x}_i,\thetab_i|\boldsymbol{d}_o,\varthetab^{(h)})+\log\phi(\boldsymbol{\theta}_i),
  \end{aligned}
\end{equation} 
where $\log f(\boldsymbol{A}_i,\boldsymbol{x}_i,\thetab_i|\d_{o},\varthetab^{(h)})=\sum_{t=1}^{T_i}\log f(\boldsymbol{A}_{it}|\boldsymbol{x}_{it},\thetab_{i};\varthetab^{(h)})$,
$\d_o=\{(\A_i,\x_i) \}_{i=1}^N$ denotes the observed data, $\d_c=\{(\A_i,\x_i,\thetab_i) \}_{i=1}^N$ denotes the complete data, and $\varthetab^{(h)}$ denotes the estimated value of
$\varthetab$ at the $h$-th iteration. Then, the expectation of the negative complete log-likelihood is 
\begin{equation}
  \label{eq:Q-function}
  \begin{aligned}
    Q(\varthetab|\varthetab^{(h)})
    &=\E[-\log(L_c(\d_{c};\varthetab|\boldsymbol{d}_o,\varthetab^{(h)}))] \\ 
    &=-\sum_{i=1}^N\int[\log f(\boldsymbol{A}_i|\boldsymbol{x}_i,\boldsymbol{\theta}_i;\varthetab)+\log\phi(\boldsymbol{\theta}_i)]\phi(\boldsymbol{\theta}_i|\boldsymbol{d}_{i,o};\varthetab^{(h)})\,d\boldsymbol{\theta}_{i}\\\
    &=-\sum_{i=1}^N\int \log
      f(\boldsymbol{A}_i|\boldsymbol{x}_i,\boldsymbol{\theta}_i;\varthetab)\phi(\boldsymbol{\theta}_i|\boldsymbol{d}_{i,o};\varthetab^{(h)})\,d\boldsymbol{\theta}_{i}\
      - \sum_{i=1}^N\int \log \phi(\boldsymbol{\theta}_i)\phi(\boldsymbol{\theta}_i|\boldsymbol{d}_{i,o};\varthetab^{(h)})\,d\boldsymbol{\theta}_{i}\ \\
    &=Q_1(\varthetab|\varthetab^{(h)})+Q_2(\varthetab^{(h)}). \\
  \end{aligned}
\end{equation}
To evaluate the integrals in $Q_1$ and $Q_2$, we propose to approximate them with Metropolis-within-Gibbs (MwG) samples
\citep{metropolis1953equation}. To do this, we draw $M$ independent MwG samples for each subject $i$.  Specifically, let
$\theta_{i,jj'}^{(h,m)}$ denote the $m$-th ($1\leq m\leq M$) simulated sample of the $(j,j')$-th element of
$\boldsymbol{\theta}_i$, given $\varthetab^{(h)}$ and $\d_o$. We draw $\theta_{i,jj'}^{(h,m)}$ from
$\mathcal{N}(\mu_{ijj'}^{(h)}, \sigma_{u,jj'}^2)$, where
$\mu_{ijj'}^{(h)}=\mathbb{E}(\theta_{i,jj'}|\boldsymbol{d}_o;\boldsymbol{\vartheta}^{(h)})$ and $\sigma^2_{u,jj'}$ is
the $(j,j')$-th element of $\boldsymbol{\Sigma}_u$. Here, $\boldsymbol{\Sigma}_u$ is a user-specified matrix, which can
be obtained by fitting element-wise GLMMs to the
data.  
In practice, we choose to use $M=100$ to approximate the integrals. Then, the Q-functions can be approximated by
\begin{equation}
  \begin{split}
    \label{eq:MwG}
    Q_1(\varthetab|\varthetab^{(h)})&=-\frac{1}{M}\sum_{m=1}^M\sum_{i=1}^N \log f(\boldsymbol{A}_i|\boldsymbol{x}_i,\boldsymbol{\theta}_i^{(h,m)};\varthetab), \\
    Q_2(\varthetab^{(h)})&=-\frac{1}{M}\sum_{m=1}^M\sum_{i=1}^N \log \phi(\boldsymbol{\theta}_i^{(h,m)}). \\  
  \end{split}
\end{equation}

In the M-step, we aim to solve
\begin{equation}
  \label{eq:M-step}
  \min_{\varthetab} ~ Q_1(\varthetab|\varthetab^{(h)})+ \gamma\fnorm{\boldsymbol{U}^{\top}\boldsymbol{U}-\boldsymbol{V}^{\top}\boldsymbol{V}}^2, \text{ subject to } \zeronorm{\Bcalb} \le sn^2.
\end{equation}
We propose to use the proximal gradient descent algorithm \citep{parikh2014} to solve it. Let
$F(\varthetab)=Q_1(\varthetab|\varthetab^{(h)})+\gamma\fnorm{\boldsymbol{U}^{\top}\boldsymbol{U}-\boldsymbol{V}^{\top}\boldsymbol{V}}^2$. In
the $h_{\kappa}$-th iteration for solving (\ref{eq:M-step}), we majorize $F(\varthetab)$ by a local quadratic function
$\mathcal{Q}_c(\varthetab)$. That
is, 
\begin{equation*}
  F(\varthetab) \le \mathcal{Q}_c(\varthetab)=F(\varthetab^{(h_{\kappa-1})})+[\nabla F(\varthetab^{(h_{\kappa-1})})](\varthetab-\varthetab^{(h_{\kappa-1})})
  +(c/2)(\varthetab-\varthetab^{(h_{\kappa-1})})^{\top}(\varthetab-\varthetab^{(h_{\kappa-1})}),
\end{equation*}
where $c$ is a positive
constant. 
Then, we solve $\min_{\varthetab}\mathcal{Q}_c(\varthetab)$ subject to $\zeronorm{\Bcalb}\leq sn^2$, which further
transfers to the proximal problem of
\begin{equation}
  \min_{\varthetab}~ \frac{1}{2}\|\varthetab-\{\varthetab^{(h_{\kappa-1})}-c^{-1}\nabla F(\varthetab^{(h_{\kappa-1})}) \}\|_F^2, \text{ subject to } \zeronorm{\Bcalb} \le sn^2.
\end{equation}
Its solution is given by $\varthetab^{(h_{\kappa})}=(\boldsymbol{U}^{(h_{\kappa})},\boldsymbol{V}^{(h_{\kappa})},\Bcalb^{(h_{\kappa})})$, where
\begin{equation}
  \label{eq:M-step-solution}
  \begin{split}
    \boldsymbol{U}^{(h_{\kappa})}&=\boldsymbol{U}^{(h_{\kappa-1})}-c^{-1} \nabla_{\boldsymbol{U}} F(\boldsymbol{U}^{(h_{\kappa-1})}\boldsymbol{V}^{(h_{\kappa-1})^{\top}},\Bcalb^{(h_{\kappa-1})}),\\
    \boldsymbol{V}^{(h_{\kappa})}&=\boldsymbol{V}^{(h_{\kappa-1})}-c^{-1} \nabla_{\boldsymbol{V}} F(\boldsymbol{U}^{(h_{\kappa})}\boldsymbol{V}^{(h_{\kappa-1})^{\top}},\Bcalb^{(h_{\kappa-1})}),\\
    \boldsymbol{\mathcal{B}}^{(h_{\kappa})}
                                 &
                                   = \text{Thr}(\boldsymbol{\mathcal{B}}^{(h_{\kappa-1})}-c^{-1}\nabla_{\Bcalb} F(\boldsymbol{U}^{(h_{\kappa})}\boldsymbol{V}^{(h_{\kappa})^{\top}},\Bcalb^{(h_{\kappa-1})}), s),
  \end{split}
\end{equation}
and $\text{Thr}(\Bcalb,s)$ is the hard-thresholding function defined as 
\begin{equation*}
  [\text{Thr}(\Bcalb,s)]_{jj'l} =
  \begin{cases}
    \boldsymbol{\mathcal{B}}_{jj'l} & \text{if $(j,j',l)\in \text{supp}(\boldsymbol{\mathcal{B}},s)$},\\
    0 & \text{otherwise},
  \end{cases}       
\end{equation*}
where $\text{supp}(\Bcalb, s)$ is the set of indices of $\Bcalb$ corresponding to its largest $sn^2$ absolute
values. This results in a sparse estimate of $\Bcalb$. 
We note that in (\ref{eq:M-step-solution}), closed-form
expressions for the gradients can be derived for specific GLMs. Furthermore, the learning rates in (\ref{eq:M-step-solution}) are $1/c$ for $\boldsymbol{U}$, $\boldsymbol{V}$, and $\Bcalb$. In practice, different learning rates can be employed for these parameters to improve numerical performance.
The M-step iterations can be terminated when \[
\max \left\{ 
\| \boldsymbol{U}^{(h_{\kappa})} - \boldsymbol{U}^{(h_{\kappa-1})} \|_F^2,\;
\| \boldsymbol{V}^{(h_{\kappa})} - \boldsymbol{V}^{(h_{\kappa-1})} \|_F^2,\;
\| \boldsymbol{\mathcal{B}}^{(h_{\kappa})} - \boldsymbol{\mathcal{B}}^{(h_{\kappa-1})} \|_F^2
\right\} < \varepsilon_M,
\]
where $\varepsilon_M$ is a user-specified threshold. After the M-step is completed, the updated estimates of $\varthetab$ can be used to re-draw MwG samples
for the Q-functions in (\ref{eq:MwG}) and update the E-step. Then, we iterate between the M- and E-steps until the objective function converges. 
The full details of the MCEM algorithm are  in Algorithm \ref{alg:1}.

For some applications, $\boldsymbol{\Theta}$ is known to be a symmetric matrix. In that scenario, we factorize it as $\boldsymbol{\Theta} = \boldsymbol{U}^\top \boldsymbol{\Lambda}
\boldsymbol{U}$ where $\boldsymbol{\Lambda}$ is a $r\times r$ diagonal matrix with diagonal entries equal to -1 or 1. Letting $\varthetab=(\boldsymbol{U},\Lambdab,\Bcalb)$, we propose to solve
\begin{equation}
\label{eq:M2}
  \min_{\varthetab} ~ -l(\boldsymbol{U}\Lambdab\boldsymbol{U}^{\top},\Bcalb)+ \gamma\fnorm{\boldsymbol{U}^{\top}\boldsymbol{U}-\boldsymbol{U}\boldsymbol{\Lambda}^{\top}\boldsymbol{\Lambda}\boldsymbol{U}^{\top}}^2, \text{ subject to } \zeronorm{\mathcal{B}} \le sn^2.
    \end{equation}
  Similar as (\ref{eq:M-step-solution}), Its solution at the $h_{\kappa}$-th iteration  is given by $\varthetab^{(h_{\kappa})}=(\boldsymbol{U}^{(h_{\kappa})},\Bcalb^{(h_{\kappa})})$, where 
  \begin{equation}
    \label{eq:M2-step-solution}
    \begin{split}
  \boldsymbol{U}^{(h_{\kappa})}&=\boldsymbol{U}^{(h_{\kappa-1})}-c^{-1} \nabla_{\boldsymbol{U}} F(\boldsymbol{U}^{(h_{\kappa-1})}\Lambdab\boldsymbol{U}^{(h_{\kappa-1})^{\top}},\Bcalb^{(h_{\kappa-1})}),\\
     \boldsymbol{\mathcal{B}}^{(h_{\kappa})}
                                 &
                                           = \text{Thr}(\boldsymbol{\mathcal{B}}^{(h_{\kappa-1})}-c^{-1}\nabla_{\Bcalb} F(\boldsymbol{U}^{(h_{\kappa})}\Lambdab\boldsymbol{U}^{(h_{\kappa})^{\top}},\Bcalb^{(h_{\kappa-1})}), s).
    \end{split}
  \end{equation} Accordingly, the $M$-step iterations can be terminated when
\begin{equation*}
  \max \left\{ 
\| \boldsymbol{U}^{(h_{\kappa})} - \boldsymbol{U}^{(h_{\kappa-1})} \|_F^2,\;
\| \boldsymbol{\mathcal{B}}^{(h_{\kappa})} - \boldsymbol{\mathcal{B}}^{(h_{\kappa-1})} \|_F^2
\right\} < \varepsilon_M.
\end{equation*}
The details of this alternative MCEM algorithm is described in Algorithm \ref{alg:2}. \\

\noindent
\begin{algorithm}[htbp]
\centering
\begin{minipage}{0.95\linewidth}
\raggedright
\textbf{Input:} Matrix response $\A$ and covariate tensor $\X$. \\
\textbf{Output:} $\Tilde{\boldsymbol{U}}, \Tilde{\boldsymbol{V}}$, and $\Tilde{\Bcalb}$ as solutions to (\ref{eq:opt}). \\[0.5em]

\textbf{Initialization:} \\
\quad Compute $\Bar{\boldsymbol{A}} = \frac{1}{N} \sum_{i=1}^{N} \boldsymbol{A}_{i}$ and let $SVD_r(g(\Bar{\boldsymbol{A}})) = [\Bar{\boldsymbol{U}}_0, \Bar{\boldsymbol{\Sigma}}_0,
  \Bar{\boldsymbol{V}}_0]$.  Set $\boldsymbol{U}^{(0)} = \Bar{\boldsymbol{U}}_0 \Bar{\boldsymbol{\Sigma}}_0^{1/2}$ and $\Bcalb^{(0)} = \mathbf{0}$. \\[0.5em]

  \textbf{Repeat} \\
  \quad At the $h$-th iteration, do\\
  \quad \textbf{E-step:} \\
  \quad Select the initial value
  $\displaystyle \boldsymbol{\theta}_{i}^{(h,0)} \sim \mathcal N_{d}(\mathbf 0,\mathbf I_{d}),
  $\\
  \quad For $m=1,\ldots, M$, perform Metropolis-within-Gibbs sampling as follows: \\
  \quad \quad Sub-step 1: Let $\thetab_i^{\dagger} = \boldsymbol{\theta}_{i}^{(h,m-1)} + \xib$, where $\xib$ is an
  $n\times n$ matrix with each entry from $N(0,\sigma_{\theta}^2)$, and $\sigma_{\theta}^2$ \\
  \hspace{16ex} is a user-specified value. \\
  \quad \quad Sub-step 2: With $f$ being defined as in (\ref{eq:MwG}), let \quad \quad \[ \boldsymbol{\theta}_{i}^{(h,m)} =
\begin{cases}
\boldsymbol{\theta}_{i}^{\dagger}, & \text{with probability } \min\left\{1, \frac{f(\boldsymbol{\theta}_{i}^{\dagger})}{f(\boldsymbol{\theta}_{i}^{(h,m-1)})} \right\}, \\
\boldsymbol{\theta}_{i}^{(h,m-1)}, & \text{otherwise}.
\end{cases}
\]
\quad Update $Q_1$ and $Q_2$ in (\ref{eq:MwG}) with $\{\thetab_i^{(h,m)} \}_{m=1}^M$.\\

\quad \textbf{M-step:} \\
\quad Initialize: Set $c = c^{(0)}$, a user-specified value.\\
\quad Repeat: At the $h_{\kappa}$-th iteration for $\kappa \geq 1$,\\
\quad \quad Sub-step 1: Let $\varthetab^{(h_{\kappa})}$ be defined as in (\ref{eq:M-step-solution}) with
$c=c^{(h_{\kappa -1})}$;\\
\quad \quad Sub-step 2: Break if $F(\varthetab^{(h_{\kappa})})\leq \mathcal{Q}_c(\varthetab^{(h_{\kappa-1})})$;\\
\quad \quad Sub-step 3: Otherwise, let $c = 2c$ and return to Sub-step 1.\\
\quad Until \[
\max \left\{ 
\| \boldsymbol{U}^{(h_{\kappa})} - \boldsymbol{U}^{(h_{\kappa-1})} \|_F^2,\;
\| \boldsymbol{V}^{(h_{\kappa})} - \boldsymbol{V}^{(h_{\kappa-1})} \|_F^2,\;
\| \boldsymbol{\mathcal{B}}^{(h_{\kappa})} - \boldsymbol{\mathcal{B}}^{(h_{\kappa-1})} \|_F^2
\right\} < \varepsilon_M.
\] \\[0.5em]

\textbf{Until} the objective function converges and set $\tilde{\boldsymbol{U}}$, $\tilde{\boldsymbol{V}}$, and $\tilde{\Bcalb}$ as the terminating values.
\end{minipage}
\caption{Monte Carlo Expectation Maximization Algorithm (MCEM)}
\label{alg:1}
\end{algorithm}

\begin{algorithm}[htbp]
\centering
\begin{minipage}{0.95\linewidth}
  \raggedright
  \textbf{Input:} Matrix response $\A$, covariate tensor $\X$ and $\Tilde{\boldsymbol{U}}, \Tilde{\boldsymbol{V}}$, and $\Tilde{\Bcalb}$ from Algorithm 1. \\
  \textbf{Output:} $\hat{\boldsymbol{U}}$, $\hat{\Lambdab}$ and $\hat{\Bcalb}$ as solutions to
  (\ref{eq:M2}). \\[0.5em]
  \textbf{Initialization:} \\
  \quad Let $\hat{\Lambdab}$ be a diagonal matrix with the $i$-th diagonal entry
  $\hat{\boldsymbol{\Lambda}}_{ii} = \operatorname{sign}(\Tilde{\boldsymbol{U}}_{.i}^{\top}
  \Tilde{\boldsymbol{V}}_{.i})$,
  $\boldsymbol{U}^{(0)} = \frac{\Tilde{\boldsymbol{U}} + \hat{\boldsymbol{\Lambda}} \Tilde{\boldsymbol{V}}^{\top}}{2}$ and $\Bcalb^{(0)} = \Tilde{\Bcalb}$.\\[0.5em]

\textbf{Repeat} \\
\quad At the $h$-th iteration, do\\
\quad \textbf{E-step:} \\
\quad Select the initial value 
$\displaystyle 
  \boldsymbol{\theta}_{i}^{(h,0)}
  \sim \mathcal N_{d}(\mathbf 0,\mathbf I_{d}),
$\\
\quad For $m=1,\ldots, M$, perform Metropolis-within-Gibbs sampling as follows: \\
\quad \quad Sub-step 1: Let $\thetab_i^{\dagger} = \boldsymbol{\theta}_{i}^{(h,m-1)} + \xib$, where $\xib$ is an
$n\times n$ matrix with each entry from $N(0,\sigma_{\theta}^2)$, and $\sigma_{\theta}^2$ \\
\hspace{16ex} is a user-specified value.  \\
\quad \quad Sub-step 2: With $f$ defined as in (\ref{eq:MwG}), let
\quad \quad \[
\boldsymbol{\theta}_{i}^{(h,m)} =
\begin{cases}
\boldsymbol{\theta}_{i}^{\dagger}, & \text{with probability } \min\left\{1, \frac{f(\boldsymbol{\theta}_{i}^{\dagger})}{f(\boldsymbol{\theta}_{i}^{(h,m-1)})} \right\}, \\
\boldsymbol{\theta}_{i}^{(h,m-1)}, & \text{otherwise}.
\end{cases}
\]
\quad Update $Q_1$ and $Q_2$ in (\ref{eq:MwG}) with $\{\thetab_i^{(h,m)} \}_{m=1}^M$.\\

\quad \textbf{M-step:} \\
\quad Initialize: Set $c = c^{(0)}$, a user-specified value.\\
\quad Repeat: at the $h_{\kappa}$-th iteration for $\kappa \geq 1$\\
\quad \quad Sub-step 1: Let $\varthetab^{(h_{\kappa})}$ be as defined in (\ref{eq:M2-step-solution}) with
$c=c^{(h_{\kappa -1})}$;;\\
\quad \quad Sub-step 2: Break if $F(\varthetab^{(h_{\kappa})})\leq \mathcal{Q}_c(\varthetab^{(h_{\kappa-1})})$;\\
\quad \quad Sub-step 3: Otherwise, let $c = 2c$ and return to Sub-step 1.\\

\quad Until \[
\max \left\{ 
\| \boldsymbol{U}^{(h_{\kappa})} - \boldsymbol{U}^{(h_{\kappa-1})} \|_F^2,\;
\| \boldsymbol{\mathcal{B}}^{(h_{\kappa})} - \boldsymbol{\mathcal{B}}^{(h_{\kappa-1})} \|_F^2
\right\} < \varepsilon_M.
\] \\[0.5em]

\textbf{Until} the objective function converges and set $\hat{\boldsymbol{U}}$ and $\hat{\Bcalb}$ as the terminating values.
\end{minipage}
\caption{Monte Carlo Expectation Maximization Algorithm with Symmetrization}
\label{alg:2}
\end{algorithm}

\subsection{Tuning Parameter Selections}
Our algorithm contains two tuning parameters $r$ and $s$ that require tuning to achieve good numerical performance. We
propose a computationally efficient two-stage approach to select the optimal tuning parameters. In the first stage, we fix
$s=0$ and choose the optimal integer of $r$ that minimizes the Extended Bayesian Information Criterion
(EBIC)\citep{chen2008ebic}, defined as
\begin{equation}
  \label{eq:ebic}
    EBIC=-2\ell(\hat{\varthetab}) + [\log(n^2N)+ C \log\{n^2(p+1)\}] \times (2nr + spn^2),
\end{equation}
where $\ell(\hat{\varthetab})$ is as defined in (\ref{eq:obs-likelihood}) and $C$ is a user-specified constant that is
set to ${1}/{2}$ by us. In the second stage, we fix $r$ at the optimal value chosen in the first stage and perform a grid
search of $s$ that minimizes the EBIC defined in (\ref{eq:ebic}). That results in the optimal combination of $r$ and $s$
being used in our proposed method. 

\section{Simulation}
\label{sec:simulation}
We conduct simulation studies to compare the numerical performance of our proposed method, named as MR-GLMM, with two
element-wise penalized generalized linear mixed models. In these two models, each entry of the response matrix is
regressed on the covariates by a generalized linear mixed model. More specifically, suppose $A_{it,jj'}$ is the
$(j,j')$-th entry of $\A_{it}$. These element-wise models assume that
\begin{equation*}
  g(A_{it,jj'})=\Theta_{jj'}+\theta_{i,jj'}+\sum_{l=1}^{p} x_{itl} \mathcal{B}_{jj'l},
\end{equation*}
where $g(\cdot)$ is the link function, $\Theta_{jj'}$, $\theta_{i,jj'}$ and $\mathcal{B}_{jj'l}$ are the $(j,j')$-th
entry of $\Thetab$, $\thetab_i$ and $\mathcal{B}_{:,:,l}$ respectively. In other words, the element-wise model can be
treated as a special case of our proposed model in (\ref{eq:link}) without considering the matrix structure. To obtain sparse estimates of $\mathcal{B}$, we solve the following penalized GLMM optimization problem for each $(j,j')$ with $1\le j,j'\le n$:
\begin{equation*}
\min_{\varthetab_{jj'}}  -l(\varthetab_{jj'}) + \nu_{jj'} \sum_{l=1}^{p} p(\mathcal{B}_{jj'l}),
\end{equation*}
where $\varthetab_{jj'}=(\Theta_{jj'},\mathcal{B}_{jj'1},\ldots,\mathcal{B}_{jj'p})^{\top}$ is the vector of parameters,
$l(\varthetab_{jj'})$ is the element-wise observed log-likelihood function similarly as defined in
(\ref{eq:obs-likelihood}), and $p(\cdot)$ is a penalty function with tuning parameter $\nu_{jj'}$. We choose $L_1$ and SCAD
penalties and call these two element-wise models as GLMM-LASSO and GLMM-SCAD respectively. In our numerical studies, we
implement these element-wise models by using the \textbf{rpql} R package \citep{Hui2017}. The optimal tuning parameters
in these models are chosen by cross-validation.


To compare the three methods, we evaluate their performance using three metrics. For $\Thetab$, we calculate the Frobenius
norm error $\fnorm{\hat{\Thetab}- \Thetab}$. For $\mathcal{B}$, we calculate both the Frobenius norm error
$\fnorm{\hat{\mathcal{B}}- \mathcal{B}}$ and the sensitivity and specificity of $\hat{\mathcal{B}}$ given by the three
methods. The sensitivity is defined as the proportion of true nonzero entries of $\mathcal{B}$ that are correctly
estimated as nonzero, and the specificity is defined as the proportion of true zero entries of $\mathcal{B}$ that are
correctly estimated as zero. We consider the following two models
\begin{align*}
&\text{Model 1 (Linear regression): } \boldsymbol{A}_{it}=\Thetab+\thetab_i+ \sum_{l=1}^{p} x_{itl} \mathcal{B}_{:,:,l}+
  \epsilonb_{it},\\
&\text{Model 2  (Logistic regression): } \text{logit}(P( \boldsymbol{A}_{it}=1))=\Thetab+\thetab_i+ \sum_{l=1}^{p} x_{itl}
  \mathcal{B}_{:,:,l}. 
\end{align*}

We generate covariates $x_{itl}$ ($1\leq i\leq N$, $1\leq t\leq T_i$ and $1\leq l\leq p$) independently from
$\mathcal{N}(0,1)$ and $\Thetab$ as $\Thetab=\bU\bU^{\top}$, where each entry of $\bU\in \mathbb{R}^{n\times r}$ is
generated from $\mathcal{N}(0,1)$. We independently generate each entry of $\thetab_i$ ($1\leq i\leq N$) from
$\mathcal{N}(0, 4)$. For each entry of $\mathcal{B}$, we generate it from Bernoulli(2,$s$), where $s=0.1$ or $0.2$,
which renders a sparse tensor of $\mathcal{B}$ with approximately 10\% and 20\% nonzero entries equal to 2. Moreover, in
Model 1, we generate each entry of $\epsilonb_{it}$ from $\mathcal{N}(0,0.25)$. In all simulations, we set $n=30$,
$p=5$, $T_i=5$ for all $1\leq i\leq N$ and choose $r=2$ or 3. For all scenarios, we run 100 simulations to calculate the
means and standard deviations of the aforementioned three metrics.

\begin{table}[h]
\centering
\small
\begin{tabular}{|c|c|c|c|c|c|c|c|c|}
  \hline
  $(n, p)$ &$N$   & $r$ & $s$ & Method & $\fnorm{\hat{\Thetab}-\Thetab}$ & $\fnorm{\hat{\mathcal{B}}-\mathcal{B}}$ & Sensitivity & Specificity \\
  \hline
  $(30, 5)$ & 200          &  2 & 0.1 & MR-GLMM & 0.59(0.09) & 5.84(2.34) & 0.99(0.01) & 1.00(0.00) \\
           &                &    &     & GLMM-LASSO & 2.94(0.83) & 37.78(0.16) & 0.22(0.01) & 0.30(0.01) \\
           &                &    &     & GLMM-SCAD & 2.91(0.81) & 37.11(0.25) & 0.24(0.01) & 0.29(0.01) \\
  \hline
           &                &    & 0.2 & MR-GLMM & 0.70(0.08) & 6.63(3.01) & 1.00(0.00) & 1.00(0.00) \\
           &                &    &     & GLMM-LASSO & 4.17(1.21) & 53.34(0.19) & 0.25(0.01) & 0.53(0.01) \\
           &                &    &     & GLMM-SCAD & 3.91(1.04) & 48.13(0.45) & 0.36(0.01) & 0.45(0.01) \\
  \hline
           &                &  3 & 0.1 & MR-GLMM & 1.22(0.29) & 9.08(2.57) & 0.97(0.02) & 1.00(0.00) \\
           &                &    &     & GLMM-LASSO & 3.33(0.59) & 37.97(0.09) & 0.23(0.01) & 0.34(0.01) \\
           &                &    &     & GLMM-SCAD & 3.20(0.54) & 35.46(0.38) & 0.31(0.02) & 0.30(0.01) \\
  \hline
           &                &    & 0.2 & MR-GLMM & 1.22(0.22) & 9.48(3.17) & 0.99(0.01) & 1.00(0.00) \\
           &                &    &     & GLMM-LASSO & 4.44(1.27) & 53.30(0.17) & 0.25(0.01) & 0.53(0.01) \\
           &                &    &     & GLMM-SCAD & 4.14(1.12) & 48.15(0.38) & 0.36(0.01) & 0.46(0.01) \\
  \hline
           & 400         &  2 & 0.1 & MR-GLMM & 0.94(0.12) & 3.48(1.81) & 0.99(0.01) & 1.00(0.00) \\
           &                &    &     & GLMM-LASSO & 2.47(0.59) & 37.80(0.13) & 0.22(0.01) & 0.30(0.01) \\
           &                &    &     & GLMM-SCAD & 2.45(0.58) & 37.38(0.21) & 0.23(0.01) & 0.29(0.01) \\
  \hline
           &                &    & 0.2 & MR-GLMM & 0.83(0.15) & 4.01(1.77) & 0.98(0.01) & 1.00(0.00) \\
           &                &    &     & GLMM-LASSO & 3.34(0.68) & 53.36(0.16) & 0.25(0.01) & 0.51(0.01) \\
           &                &    &     & GLMM-SCAD & 3.18(0.60) & 48.43(0.37) & 0.35(0.01) & 0.44(0.01) \\
  \hline
           &                &  3 & 0.1 & MR-GLMM & 0.86(0.17) & 6.57(2.70) & 0.99(0.01) & 1.00(0.00) \\
           &                &    &     & GLMM-LASSO & 2.82(0.38) & 37.77(0.11) & 0.24(0.01) & 0.29(0.01) \\
           &                &    &     & GLMM-SCAD & 2.75(0.34) & 35.63(0.31) & 0.30(0.01) & 0.27(0.01) \\
  \hline
           &                &    & 0.2 & MR-GLMM & 1.48(0.38) & 6.02(2.50) & 0.96(0.02) & 1.00(0.00) \\
           &                &    &     & GLMM-LASSO & 3.35(0.73) & 53.31(0.18) & 0.25(0.01) & 0.51(0.01) \\
           &                &    &     & GLMM-SCAD & 3.15(0.63) & 48.21(0.41) & 0.36(0.01) & 0.44(0.01) \\
  \hline
\end{tabular}
\caption{Simulation results for Model 1.}
\label{tab:1}
\end{table}

\begin{table}[h]
\centering
\small
\begin{tabular}{|c|c|c|c|c|c|c|c|c|}
  \hline
  $(n,p)$ & $N$ & $r$ & $s$ & Method & $\fnorm{\hat{\Thetab}-\Thetab}$ & $\fnorm{\hat{\mathcal{B}}-\mathcal{B}}$
  & Sensitivity & Specificity \\
  \hline
$(30,5)$ & 200         &  2 & 0.1 & MR-GLMM & 2.27(3.1) & 4.12(2.22) & 0.99(0.01) & 1(0) \\
    &                &    &     & GLMM-LASSO & 12.92(1.82) & 42.28(0.01) & 0.51(0.04) & 1(0) \\
    &                &    &     & GLMM-SCAD & 12.94(1.78) & 42.29(0.01) & 0.56(0.04) & 1(0) \\
    \hline
    &                &    & 0.2 & MR-GLMM & 3.35(4.33) & 5.64(2.86) & 0.99(0.01) & 1(0) \\
    &                &    &     & GLMM-LASSO & 17.79(3.57) & 59.85(0.01) & 0.39(0.03) & 1(0) \\
    &                &    &     & GLMM-SCAD & 17.79(3.57) & 59.84(0.01) & 0.47(0.05) & 1(0) \\
    \hline
    &                &  3 & 0.1 & MR-GLMM & 2.33(2.02) & 9.18(3.25) & 0.99(0.01) & 1(0) \\
    &                &    &     & GLMM-LASSO & 15.84(1.71) & 42.3(0.02) & 0.45(0.05) & 1(0) \\
    &                &    &     & GLMM-SCAD & 15.84(1.71) & 42.3(0.01) & 0.52(0.04) & 1(0) \\
    \hline
    &                &    & 0.2 & MR-GLMM & 2.64(2.07) & 11.23(3.61) & 0.99(0.01) & 1(0) \\
    &                &    &     & GLMM-LASSO & 21.63(2.38) & 59.85(0.01) & 0.37(0.04) & 1(0) \\
    &                &    &     & GLMM-SCAD & 21.63(2.38) & 59.85(0.01) & 0.43(0.02) & 1(0) \\
  \hline
 & 400 & 2  & 0.1 & MR-GLMM & 2.03(2.38) & 3.47(2.36) & 0.99(0.02) & 1(0) \\
    &                &    &     & GLMM-LASSO & 13.58(2.02) & 42.28(0.01) & 0.59(0.03) & 1(0) \\
    &                &    &     & GLMM-SCAD & 13.58(2.02) & 42.28(0.01) & 0.62(0.03) & 1(0) \\
    \hline
    &                &    & 0.2 & MR-GLMM & 3.32(4.41) & 4.73(2.81) & 0.99(0.01) & 1(0) \\
    &                &    &     & GLMM-LASSO & 17.32(2.41) & 59.82(0.01) & 0.52(0.04) & 1(0) \\
    &                &    &     & GLMM-SCAD & 17.32(2.41) & 59.82(0.01) & 0.57(0.03) & 1(0) \\
    \hline
    &                &  3 & 0.1 & MR-GLMM & 3.52(4.03) & 5.4(2.33) & 0.99(0.01) & 1(0) \\
    &                &    &     & GLMM-LASSO & 15.45(2.01) & 42.28(0.01) & 0.58(0.04) & 1(0) \\
    &                &    &     & GLMM-SCAD & 15.46(2.04) & 42.28(0.01) & 0.62(0.03) & 1(0) \\
    \hline
    &                &    & 0.2 & MR-GLMM & 5.09(3.2) & 6.59(3.49) & 0.97(0.02) & 1(0) \\
    &                &    &     & GLMM-LASSO & 21.86(2.52) & 59.84(0.01) & 0.48(0.04) & 1(0) \\
    &                &    &     & GLMM-SCAD & 21.98(2.74) & 59.85(0.04) & 0.54(0.03) & 1(0) \\
  \hline
\end{tabular}
\caption{Simulation results for Model 2.}
\label{tab:2}
\end{table}
The simulation results are tabulated in Tables \ref{tab:1} and \ref{tab:2}. The results demonstrate that our proposed
method substantially outperforms the two alternative methods across all scenarios and metrics. For the estimation error
of $\Thetab$, our method consistently achieves much smaller errors in all cases. This superior performance arises from
our method's modeling of the low-rank structure of $\Thetab$, which the two element-wise GLMM models fail to
utilize. Regarding variable selection and estimation error of $\mathcal{B}$, our method achieves nearly perfect
selection performance, exhibiting high sensitivity and specificity across all cases. In contrast, despite employing
penalization techniques to produce sparse estimators of $\mathcal{B}$, the GLMM-LASSO and GLMM-SCAD methods exhibit
inferior variable selection performance compared to our approach, particularly in terms of sensitivity. This deficiency
again arises from their failure to leverage the matrix structure of $\mathcal{B}$. Additionally, we observe that the
estimation errors of $\mathcal{B}$ obtained by our method are substantially smaller than those produced by the two
alternatives. In summary, the simulation studies provide compelling evidence that our method possesses significant
advantages over element-wise approaches.



\section{Applications to brain imaging data}
\label{sec:appl-brain-imag}
\subsection{DTI imaging data}
\label{sec:dti-imaging-data}
We apply our method to analyze an ADNI brain imaging dataset acquired using DTI. The cohort consists of $N = 250$
healthy control participants, aged 55 to 89 years, who are at risk for Alzheimer's disease. For each subject, DTI scans
produce 100 measurements corresponding to tractography-based brain connectivity measures derived from non-overlapping
segments along white matter fiber tracts. These 100 repeated measures are grouped into 5 consecutive time windows, each
comprising 20 measurements. Within each window, we use these 20 samples to compute the sample covariance matrix among 90
ROIs, denoted as $A_{it} \in \mathbb{R}^{90 \times 90}$, which represents the structural brain network of subject $i$ at
time window $t \in \{1, \dots, 5\}$.  We hypothesize that (1) AD risk factors such as age and APOE4 carrier status may
induce alterations in the brain connectome, and (2) these detrimental effects exhibit sexual differences.

To verify our hypothesis, we regress $A_{it}$ on age, sex, APOE4 and DX status by our proposed matrix-response linear models:
\begin{align*} 
    &\text{Model 1}: \boldsymbol{A}_{it}=\boldsymbol{\Theta} + \boldsymbol{\theta}_i + \mathcal{B}_1\text{Sex}_i +\mathcal{B}_2\text{Age}_i+\mathcal{B}_3\text{Sex}_i\times\text{Age}_i + \boldsymbol{\epsilon}_{it}.\\
    &\text{Model 2}: \boldsymbol{A}_{it}=\boldsymbol{\Theta} + \boldsymbol{\theta}_i + \mathcal{B}_1\text{Sex}_i +\mathcal{B}_2\text{APOE4}_i+\mathcal{B}_3\text{Sex}_i\times\text{APOE4}_i + \boldsymbol{\epsilon}_{it}.\\
    &\text{Model 3}: \boldsymbol{A}_{it}=\boldsymbol{\Theta} + \boldsymbol{\theta}_i + \mathcal{B}_1\text{Sex}_i +\mathcal{B}_2\text{DX}_i+\mathcal{B}_3\text{Sex}_i\times\text{DX}_i + \boldsymbol{\epsilon}_{it}.
\end{align*}
In these models, female serves as the reference category for sex. APOE4 status is treated as an ordinal variable with three levels: no copies (non-carriers), one copy (heterozygous carriers), and two copies (homozygous carriers) of the APOE4 allele. Diagnostic (DX) status is also treated as an ordinal variable representing AD progression, ranging from Cognitively Normal (CN), Subjective Memory Complaints (SMC), Early/Late Mild Cognitive Impairment (EMCI/LMCI), to Alzheimer's Disease (AD).

For each model, we initialize our algorithm by running an elementwise linear mixed model and using the resulting estimators
as the initial estimators of $\mathcal{B}_1,\mathcal{B}_2$ and $\mathcal{B}_3$. The optimal values of tuning parameters are
selected by the EBIC within the range that $r \in \{1, 2, 3, 4\}$ and $s \in \{0.01, 0.02, 0.05, 0.08, 0.1\}$. To visualize
$\hat{\boldsymbol{\Theta}}$ obtained by our method, we present heatmaps of $\hat{\boldsymbol{\Theta}}$ for each model and
apply $K$-means clustering to the rows and columns of $\hat{\boldsymbol{\Theta}}$. The optimal number of clusters is
determined using the Silhouette method  \citep{rousseeuw1987silhouettes}, which evaluates clustering quality by measuring how
similar objects are to their assigned cluster compared to other clusters, through a combination of intra-cluster cohesion
and inter-cluster separation metrics. We then identify regions from the heatmaps that correspond to the eight established
brain functional modules: Visual, Sensorimotor, Dorsal Attention, Ventral Attention, Limbic, Frontoparietal, Default Mode,
and Cerebellum modules. These modules represent distinct neural networks associated with various cognitive and sensory
functions, providing a structured framework for interpreting the patterns observed in our analysis. To visualize
$\hat{\mathcal{B}}_j$ obtained by our method, we plot the axial and sagittal views of the networks identified by
$\hat{\mathcal{B}}_j$. Specifically, for each nonzero entry in $\hat{\mathcal{B}}_j$, we draw an edge between the two ROIs
corresponding to that entry. To simplify the presentation of these networks, all ROIs are grouped into the eight functional
modules. 

\begin{figure}[hbtp]
    \centering
    \begin{minipage}{0.4\textwidth}
        \centering
        \includegraphics[scale=0.12]{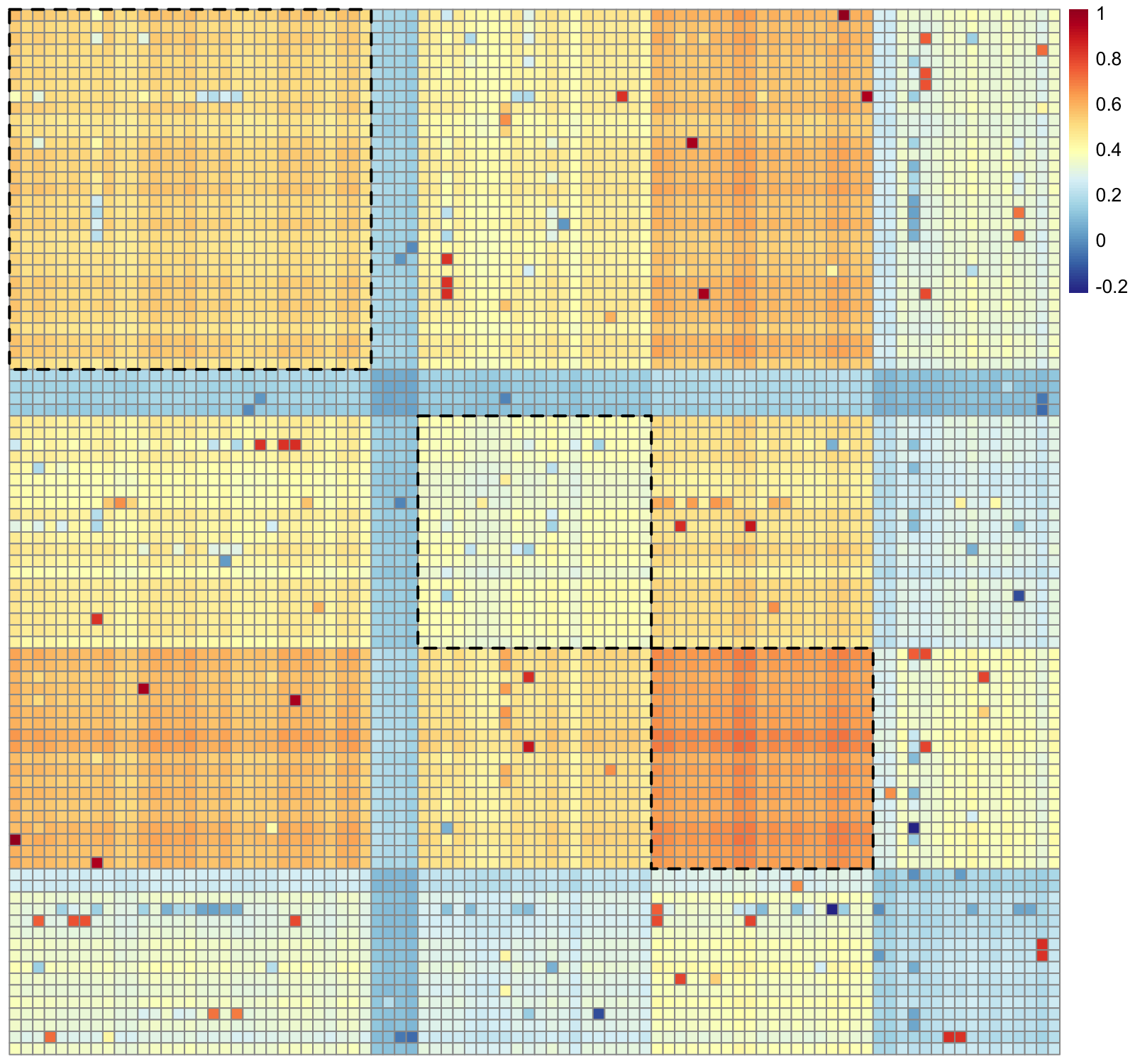}
    \end{minipage}%
    \begin{minipage}{0.5\textwidth}  
        \centering
        \resizebox{0.8\textwidth}{!}{  
            \begin{tabular}{c c}
                \hline
                Module & Large-scale functional networks \\
                \hline
                1 & Default Mode, Visual, Frontoparietal, Sensorimotor, Ventral Attention \\
                2 & Default Mode, Limbic, Visual, Sensorimotor \\
                3 & Default Mode, Limbic, Visual, Sensorimotor, Dorsal Attention \\
                \hline
            \end{tabular}
        }
        \end{minipage}
        \caption{The heatmap of $\hat{\boldsymbol{\Theta}}$ in Model 1, where rows and columns are arranged according to the
          \textit{K}-means clustering. The black dashed lines represent the groups of functional modules identified in the
          clustering process with module names on the right table.}
    \label{fig:model1-theta}
\end{figure}

\begin{figure}[t]
\centering

\begin{minipage}[t]{0.76\textwidth}
    \centering
    \includegraphics[width=\linewidth]{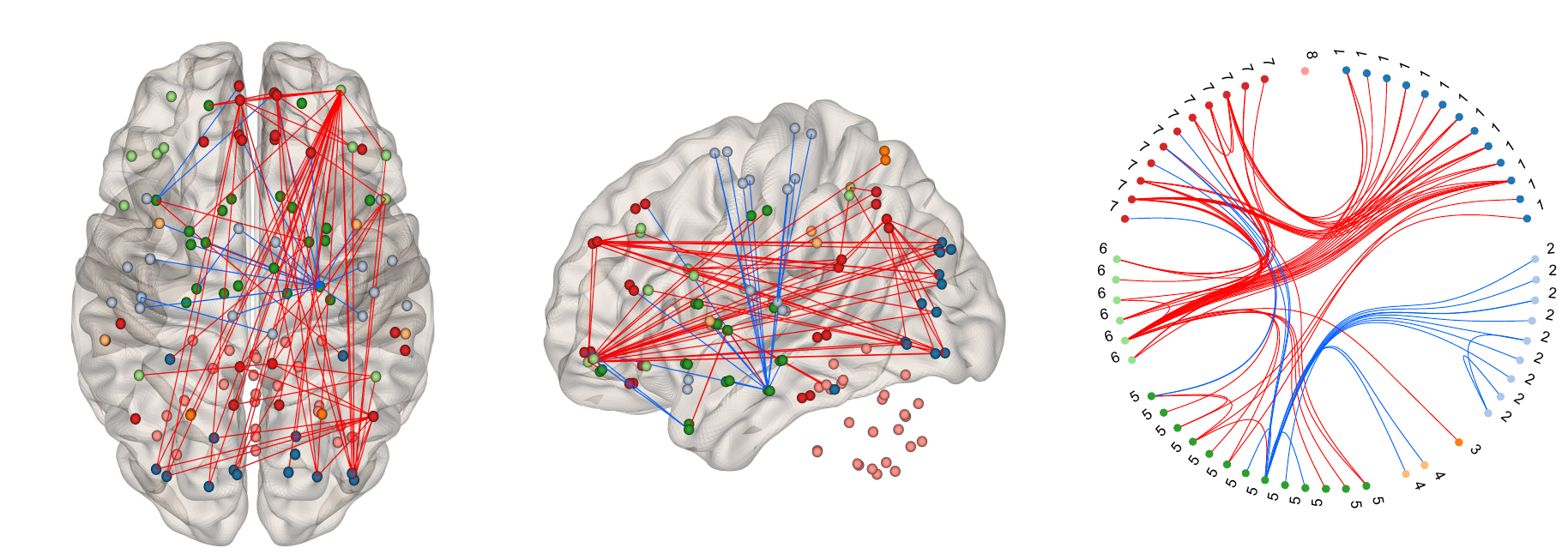}
\end{minipage}%
\hfill
\begin{minipage}[t]{0.22\textwidth}
    \centering
    \includegraphics[width=0.6\linewidth]{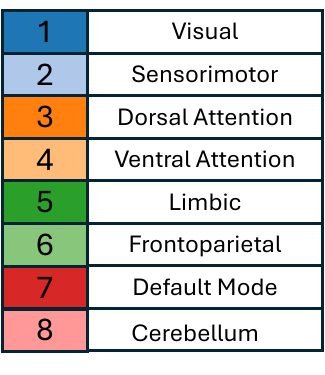}
\end{minipage}

\vspace{1ex}

\noindent
\textit{(a) Axial and sagittal views of $\hat{\mathcal{B}}_2$ and the corresponding functional region network.}

\vspace{2ex}

\begin{minipage}[t]{0.76\textwidth}
    \centering
    \includegraphics[width=\linewidth]{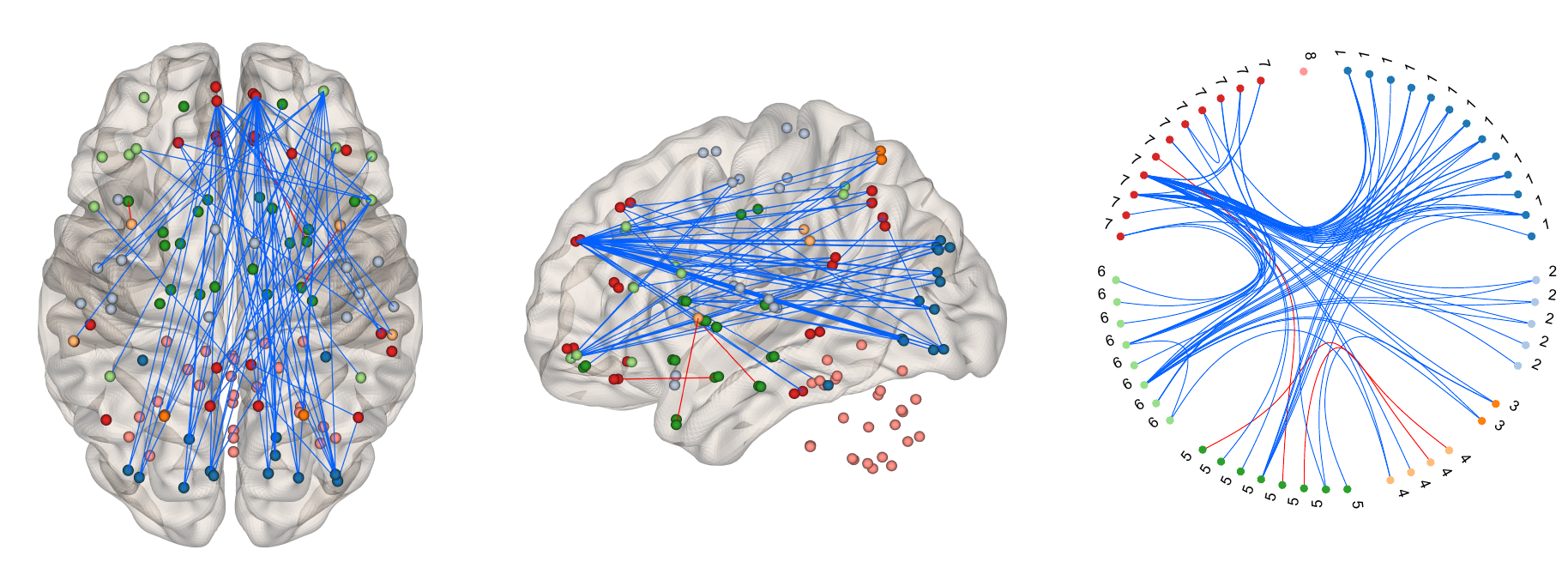}
\end{minipage}%
\hfill
\begin{minipage}[t]{0.22\textwidth}
    \centering
    \includegraphics[width=0.6\linewidth]{legend_AAL.pdf}
\end{minipage}

\vspace{1ex}

\noindent
\textit{(b) Axial and sagittal views of $\hat{\mathcal{B}}_3$ and the corresponding functional region network.}

\caption{Estimated $\hat{\mathcal{B}}_2$ and $\hat{\mathcal{B}}_3$ in Model 1. Red and blue edges indicate positive and negative entries in $\hat{\mathcal{B}}_2$ and $\hat{\mathcal{B}}_3$, respectively.}
\label{fig:model1-b2b3}
\end{figure}

The results for Model 1 are shown in Figures \ref{fig:model1-theta} and
  \ref{fig:model1-b2b3}. 
In Figure \ref{fig:model1-theta}, Module 1 exhibits positive correlations among the Default Mode network (DMN), Visual network (VN), Frontoparietal network (FPN), Sensorimotor network (SMN), and Ventral Attention network (VAN), suggesting coordinated activity in cognitive, sensory, and attentional processing. 
Module 2 demonstrates connectivity patterns linking the DMN, Limbic network (LN), and VN, and SMN, elucidating the integration between higher-order cognitive functions and affective processing systems. Module 3
  expands this pattern, incorporating the Dorsal Attention network (DAN), which indicates a broader integration of attentional
  control. Notably, DMN and VN remain consistently correlated across all groups, highlighting
  their fundamental role in intrinsic brain connectivity. Figure \ref{fig:model1-b2b3}(a) shows that with increasing age,
  the VN, LN, FPN, and DMN exhibit stronger positive inter-correlations, suggesting
  enhanced connections among these large-scale cognitive and associative networks with aging. On the other hand, the connections between LN, SMN, and DMN are reduced, indicating a potential shift
  in emotional regulation and sensorimotor processing with aging. The age and sex interactions depicted in Figure
  \ref{fig:model1-b2b3}(b) reveals that older males demonstrate reduced inter-module correlations across all major
  networks—with the exception of the Cerebellum—when compared to younger females. This pattern suggests that large-scale
  brain network connectivity undergoes age- and sex-dependent modulation, potentially reflecting differences in
  neuroplasticity and the adaptation of functional interactions throughout the lifespan. 

\begin{figure}[hbtp]
    \centering
    \begin{minipage}{0.4\textwidth}
        \centering
        \includegraphics[scale=0.12]{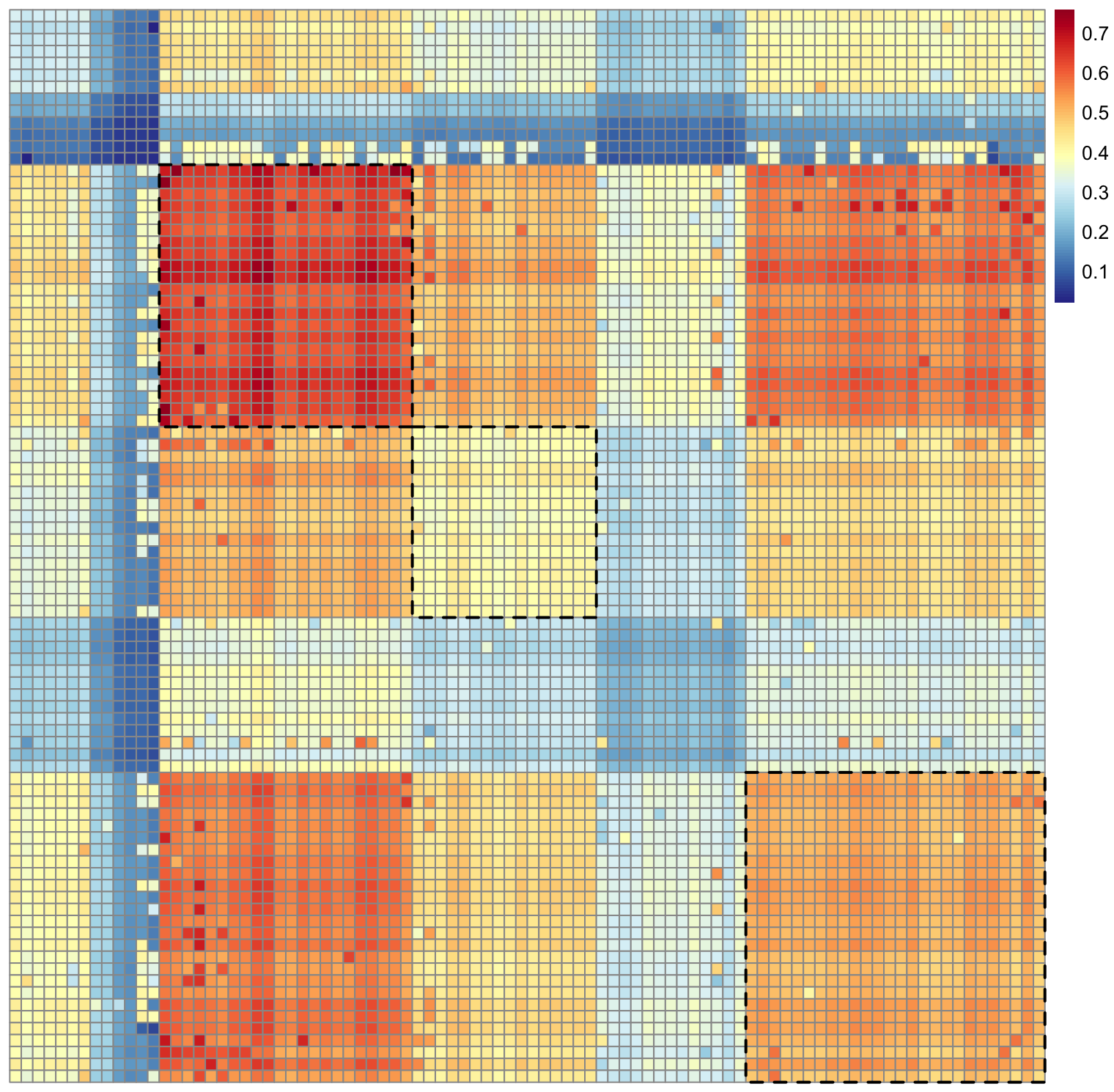}
    \end{minipage}%
    \begin{minipage}{0.5\textwidth}  
        \centering
        \resizebox{0.8\textwidth}{!}{  
            \begin{tabular}{c c}
                \hline
                Module & Large-scale functional networks \\
                \hline
                1 & Default Mode, Limbic, Visual, Frontoparietal, Sensorimotor, Dorsal Attention \\
                2 & Default Mode,Limbic, Visual, Frontoparietal, Sensorimotor\\
                3 & Default Mode, Visual, Sensorimotor, Ventral Attention \\
                \hline
            \end{tabular}
        }
        \end{minipage}
        \caption{The heatmap of $\hat{\boldsymbol{\Theta}}$ in Model 2, where rows and columns are arranged according to the
          \textit{K}-means clustering. The black dashed lines represent the groups of functional modules identified in the
          clustering process with module names on the right table.}
    \label{fig:model2-theta}
\end{figure}

\begin{figure}[t]
\centering

\begin{minipage}[t]{0.76\textwidth}
    \centering
    \includegraphics[width=\linewidth]{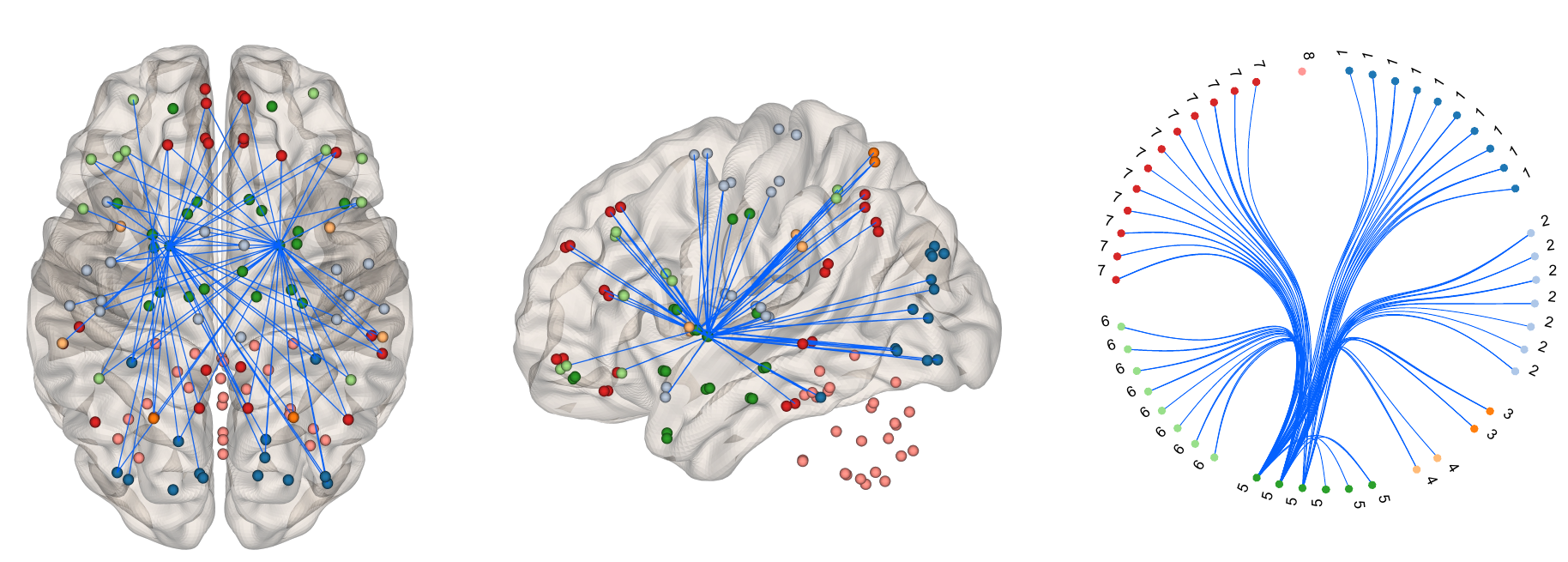}
\end{minipage}%
\hfill
\begin{minipage}[t]{0.18\textwidth}
    \centering
    \includegraphics[width=0.6\linewidth]{legend_AAL.pdf}
\end{minipage}

\vspace{1ex}

\noindent
\textit{(a) Axial and sagittal views of $\hat{\mathcal{B}}_2$ and the corresponding functional region network.}

\vspace{2ex}

\begin{minipage}[t]{0.76\textwidth}
    \centering
    \includegraphics[width=\linewidth]{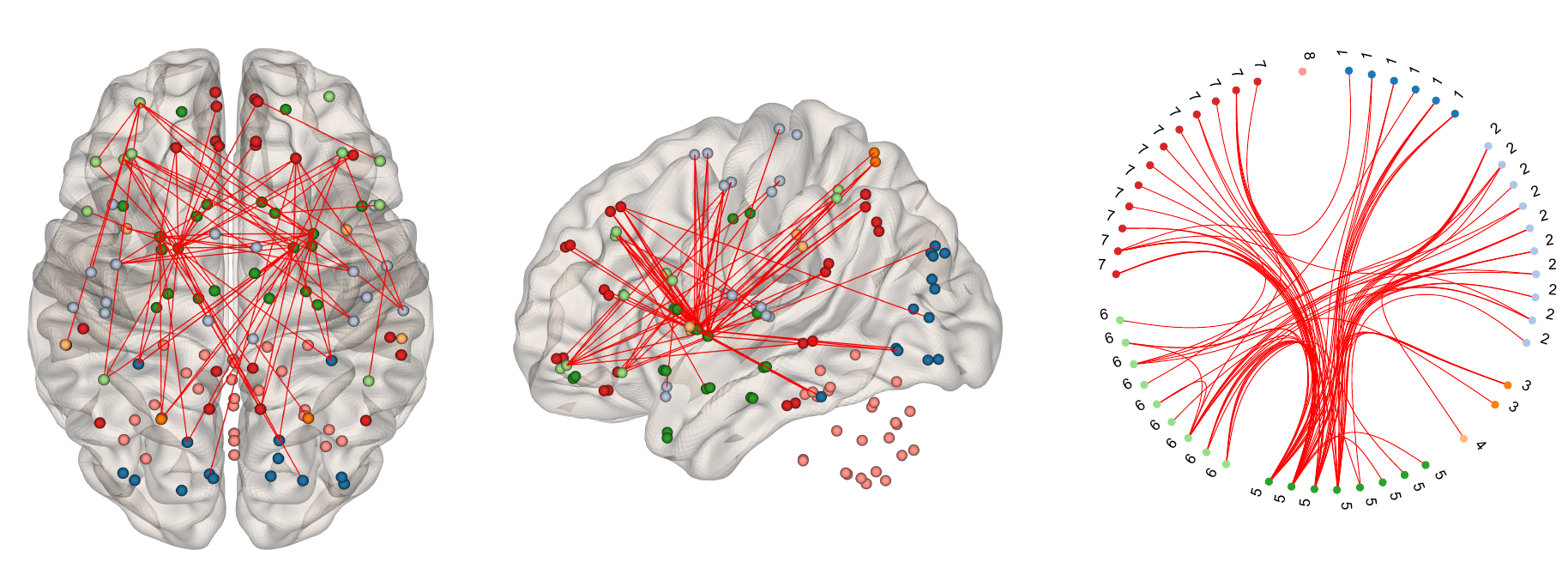}
\end{minipage}%
\hfill
\begin{minipage}[t]{0.18\textwidth}
    \centering
    \includegraphics[width=0.6\linewidth]{legend_AAL.pdf}
\end{minipage}

\vspace{1ex}

\noindent
\textit{(b) Axial and sagittal views of $\hat{\mathcal{B}}_3$ and the corresponding functional region network.}

\caption{Estimated $\hat{\mathcal{B}}_2$ and $\hat{\mathcal{B}}_3$ in Model 2. Red and blue edges indicate positive and negative entries in $\hat{\mathcal{B}}_2$ and $\hat{\mathcal{B}}_3$, respectively.}
\label{fig:model2-b2b3}
\end{figure}

The results for Model 2 are shown in Figures \ref{fig:model2-theta} and \ref{fig:model2-b2b3}. In Figure
\ref{fig:model2-theta}, Module 1 exhibits extensive connectivity across multiple networks—DMN, LN, VN,
FPN, SMN, and DAN—suggesting robust integration between cognitive, emotional, and
attentional systems. Module 2 displays a similar connectivity profile but excludes the DAN, indicating
a more focused functional integration primarily involving cognitive processes, sensory integration, and emotional
regulation. Module 3 shows a more selective connectivity pattern, connecting only the DMN, VN, SMN,
and VAN, potentially reflecting a greater emphasis on externally-oriented attentional
processing. Notably, despite these regional differences, the DMN, VN, and SMN maintain
positive correlations across all regions, underscoring their fundamental importance in the brain's intrinsic functional architecture. 
Figure \ref{fig:model2-b2b3}(a) reveals that APOE4 carriers exhibit progressively stronger negative
inter-correlations between LN and VN, SMN, FPN, and DMN. Additionally, both DN and VAN in APOE4 carriers display negative associations with
the Limbic system, suggesting widespread disruption in functional integration across multiple cognitive and sensorimotor
domains. We also observe negative correlations between SMN and both LN and DMN, potentially reflecting APOE4-associated disruptions in the coordination of motor functions, emotional processing,
and internally directed cognitive processes. Regarding sex and APOE4 interactions shown in Figure \ref{fig:model2-b2b3},
male participants with elevated APOE4 levels demonstrate stronger inter-module connectivity across all major networks except
the Cerebellum, compared to both males with lower APOE4 levels and females regardless of APOE4 status. This pattern suggests
sex-specific modulation of large-scale brain network organization, possibly reflecting compensatory neural mechanisms or
differential vulnerability to APOE4-associated neurobiological changes.

\begin{figure}[hbtp]
    \centering
    \begin{minipage}{0.4\textwidth}
        \centering
        \includegraphics[scale=0.12]{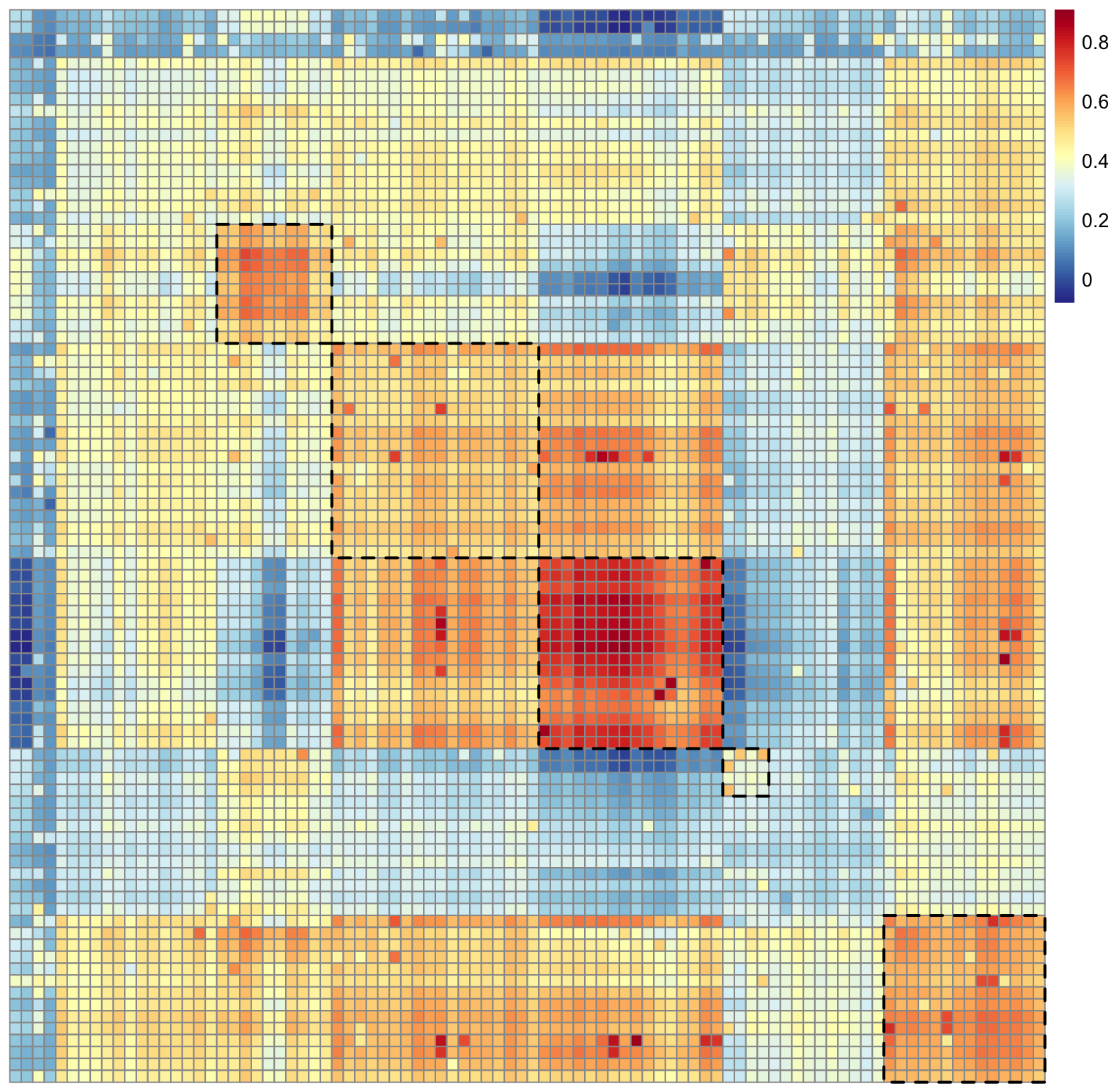}
    \end{minipage}%
    \begin{minipage}{0.5\textwidth}  
        \centering
        \resizebox{0.9\textwidth}{!}{  
            \begin{tabular}{c c}
                \hline
                Module & Large-scale functional networks \\
                \hline
                1      & Default Mode \\
                2      & Frontoparietal, Sensorimotor, Ventral Attention \\
                3      & Visual \\
                4      & Limbic \\
                5      & Default Mode, Frontoparietal, Sensorimotor \\
                \hline
            \end{tabular}
        }
        \end{minipage}
        \caption{The heatmap of $\hat{\boldsymbol{\Theta}}$ in Model 3, where rows and columns are arranged according to the
          \textit{K}-means clustering. The black dashed lines represent the groups of functional modules identified in the
          clustering process with module names on the right table.}
    \label{fig:model3-theta}
\end{figure}

\begin{figure}[t]
\centering

\begin{minipage}[t]{0.76\textwidth}
    \centering
    \includegraphics[width=\linewidth]{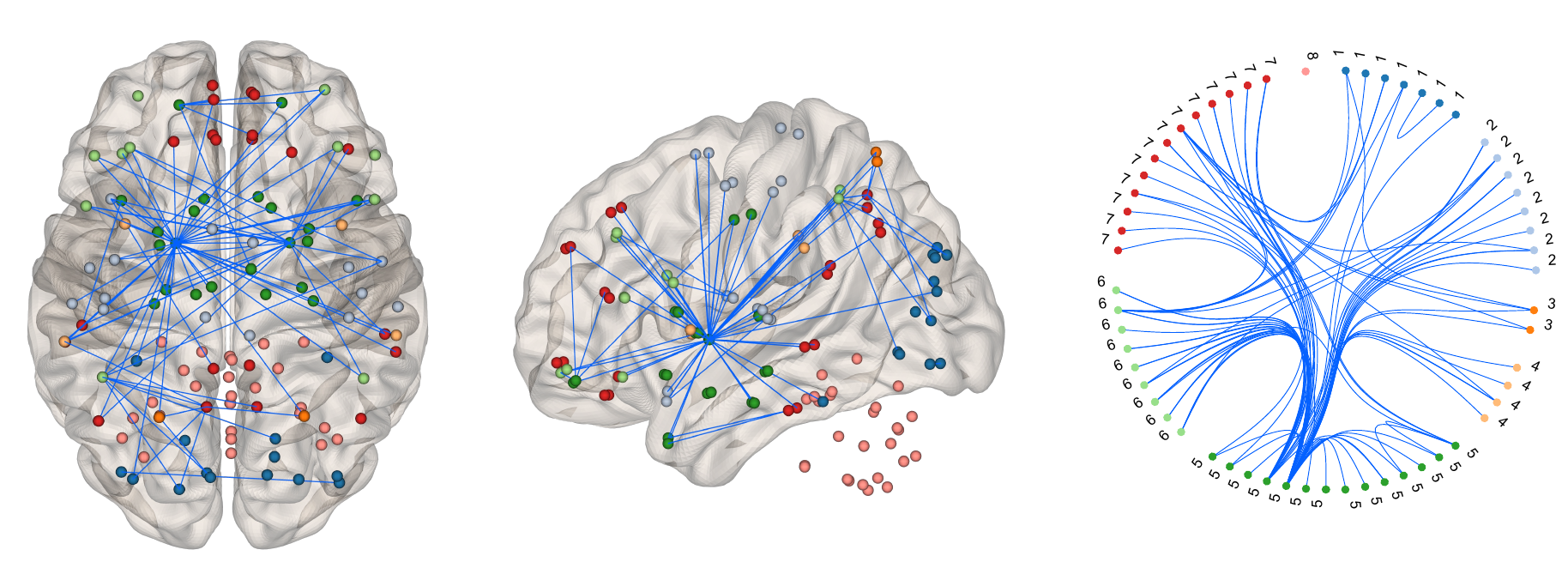}
\end{minipage}%
\hfill
\begin{minipage}[t]{0.18\textwidth}
    \centering
    \includegraphics[width=0.6\linewidth]{legend_AAL.pdf}
\end{minipage}

\vspace{1ex}

\noindent
\textit{(a) Axial and sagittal views of $\hat{\mathcal{B}}_2$ and the corresponding functional region network.}

\vspace{2ex}

\begin{minipage}[t]{0.76\textwidth}
    \centering
    \includegraphics[width=\linewidth]{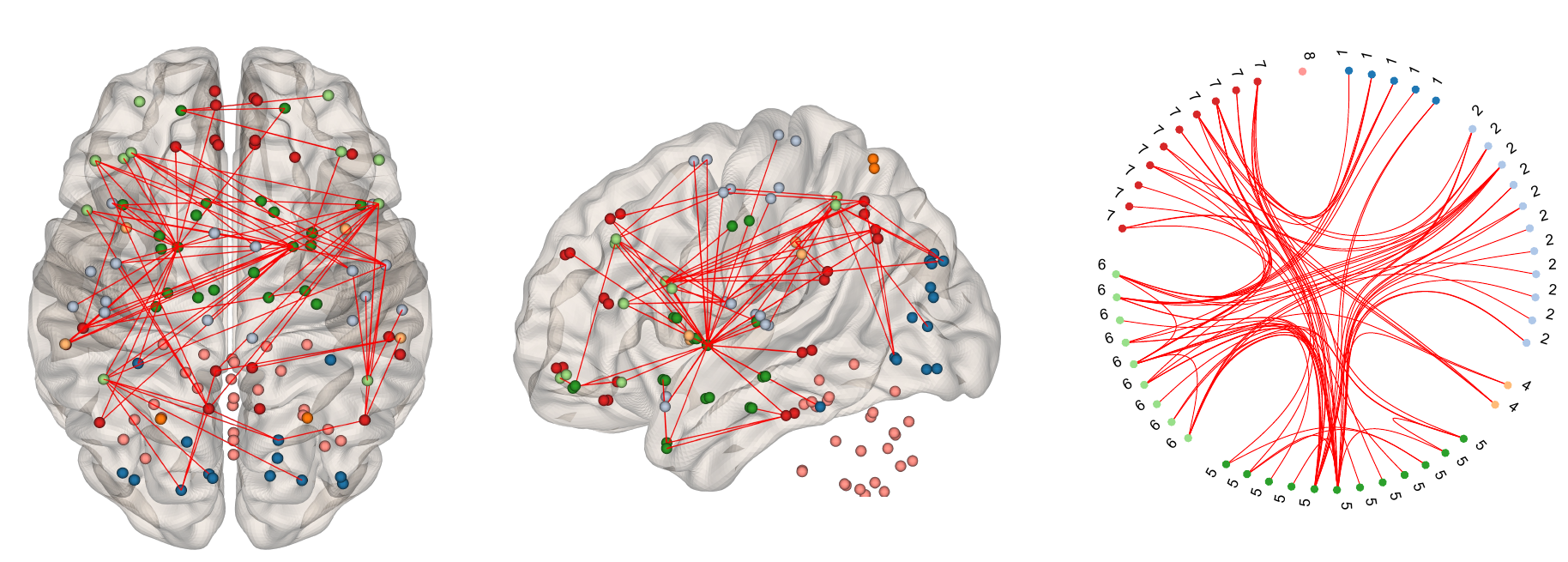}
\end{minipage}%
\hfill
\begin{minipage}[t]{0.18\textwidth}
    \centering
    \includegraphics[width=0.6\linewidth]{legend_AAL.pdf}
\end{minipage}

\vspace{1ex}

\noindent
\textit{(b) Axial and sagittal views of $\hat{\mathcal{B}}_3$ and the corresponding functional region network.}

\caption{Estimated $\hat{\mathcal{B}}_2$ and $\hat{\mathcal{B}}_3$ in Model 3. Red and blue edges indicate positive and negative entries in $\hat{\mathcal{B}}_2$ and $\hat{\mathcal{B}}_3$, respectively.}
\label{fig:model3-b2b3}
\end{figure}

The results for Model 3 are illustrated in Figures \ref{fig:model3-theta} and \ref{fig:model3-b2b3}. Figure
\ref{fig:model3-theta} shows that Module 1 is contained in DMN, which is known to mediate
self-referential thought and introspection. Module 3 is completely encompassed by the VN, representing areas
involved in primary and secondary visual processing. Likewise, Module 4 is entirely contained within LN,
which is crucial for emotion regulation and memory formation. In contrast, Modules 2 and 5 comprise nodes from multiple large-scale networks—FPN, SMN, VAN, and DMN—suggesting their involvement in complex cognitive
functions, attentional processes, and motor coordination. Figure \ref{fig:model3-b2b3}(a) illustrates that Alzheimer's disease progression, characterized by symptoms of memory impairment and cognitive decline, is associated with progressively stronger negative inter-correlations between LN and several other networks, including VN, SMN, FPN, and DMN. Additionally, specific regions within both DN and VAN show negative associations with the Limbic system. These patterns suggest widespread disruption in functional integration across multiple cognitive and sensorimotor domains as AD severity increases. Conversely, Figure \ref{fig:model3-b2b3}(b) shows that males with escalated AD progression exhibit stronger inter-module connectivity throughout nearly all major networks, with the exception of the Cerebellum. This enhanced connectivity in males with elevated AD progression is evident when compared both to males at early stage and to females across all AD stages.

These findings align with established sex-specific differences in large-scale brain networks. Previous
  research has documented that males typically exhibit stronger right-lateralized connectivity patterns
   \citep{tomasi2012lateralization}, while females demonstrate greater interhemispheric integration, potentially contributing
  to distinct compensatory mechanisms during neurodegenerative progression. Furthermore, neurodegenerative diseases
  characteristically target specific functional modules rather than causing uniform atrophy, which prompts compensatory
  reorganization within brain networks  \citep{ingalhalikar2014sex}. The observed increase in inter-module connectivity among
  males with elevated disease status likely represents an adaptive response aimed at preserving functional stability despite
  ongoing network disruptions. 


\subsection{Functional MRI Data}
\begin{figure}[hbtp]
    \centering
    \begin{minipage}{0.4\textwidth}
        \centering
        \includegraphics[scale=0.12]{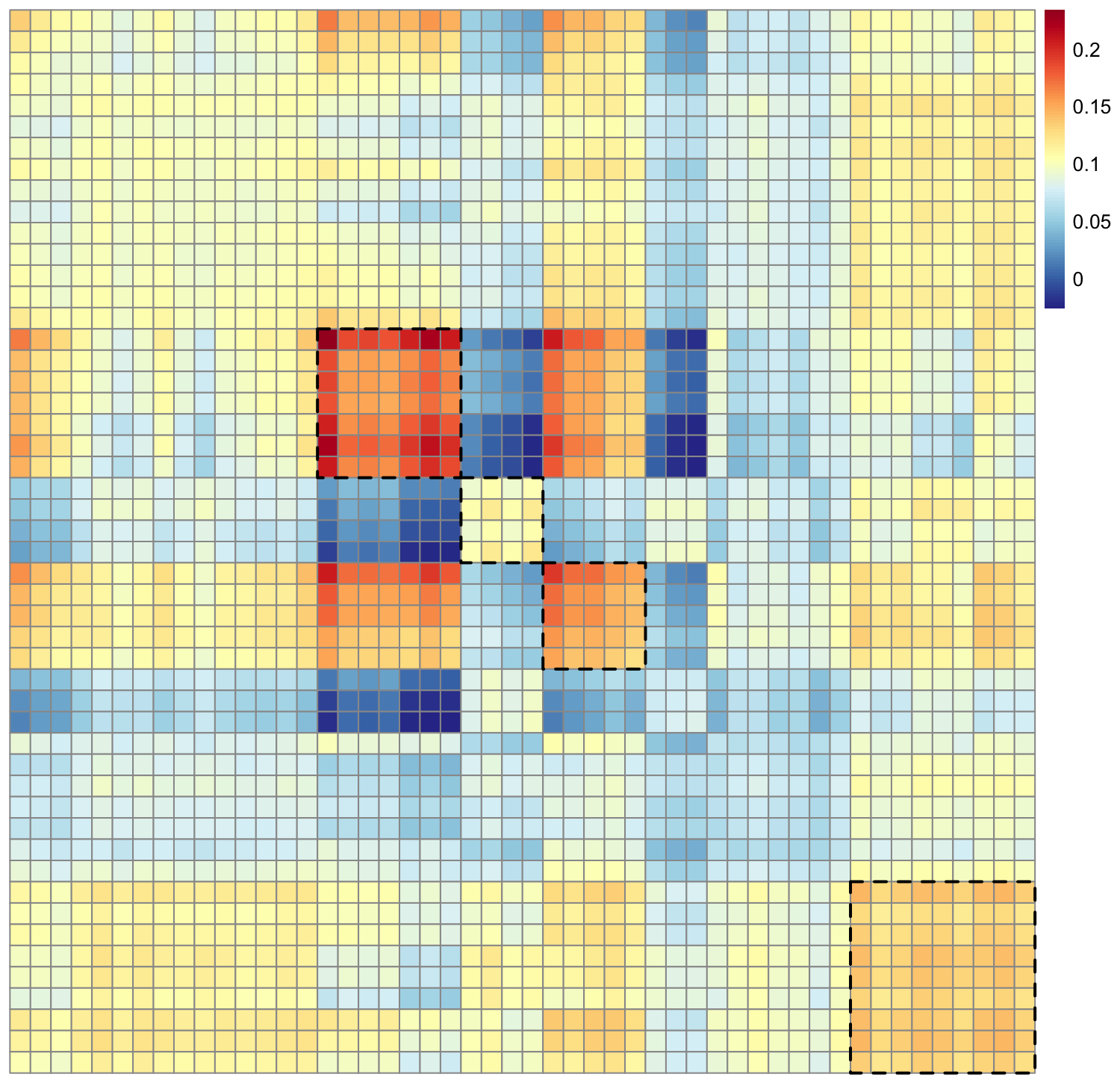}
    \end{minipage}%
    \begin{minipage}{0.5\textwidth}  
        \centering
        \resizebox{\textwidth}{!}{  
            \begin{tabular}{c c}
                \hline
                Module & Large-scale functional networks \\
                \hline
                1      & Early Visual, Ventral Stream Visual, Dorsal Stream Visual, MT+ Complex and Neighboring Visual Areas \\
                2      & Posterior Cingulate \\
                3      & Early Visual, Dorsal Stream Visual, Superior Parietal\\
                4      & Somatosensory and Motor, Paracentral Lobular and Mid Cingulate \\
                \hline
            \end{tabular}
        }
        \end{minipage}
        \caption{The heatmap of $\hat{\boldsymbol{\Theta}}$, where rows and columns are arranged according to the
          \textit{K}-means clustering. The black dashed lines represent the groups of functional modules identified in the
          clustering process with module names on the right table.}
    \label{fig:theta}
\end{figure}

\begin{figure}[h]
\centering

\begin{minipage}[t]{0.76\textwidth}
    \centering
    \includegraphics[width=\linewidth]{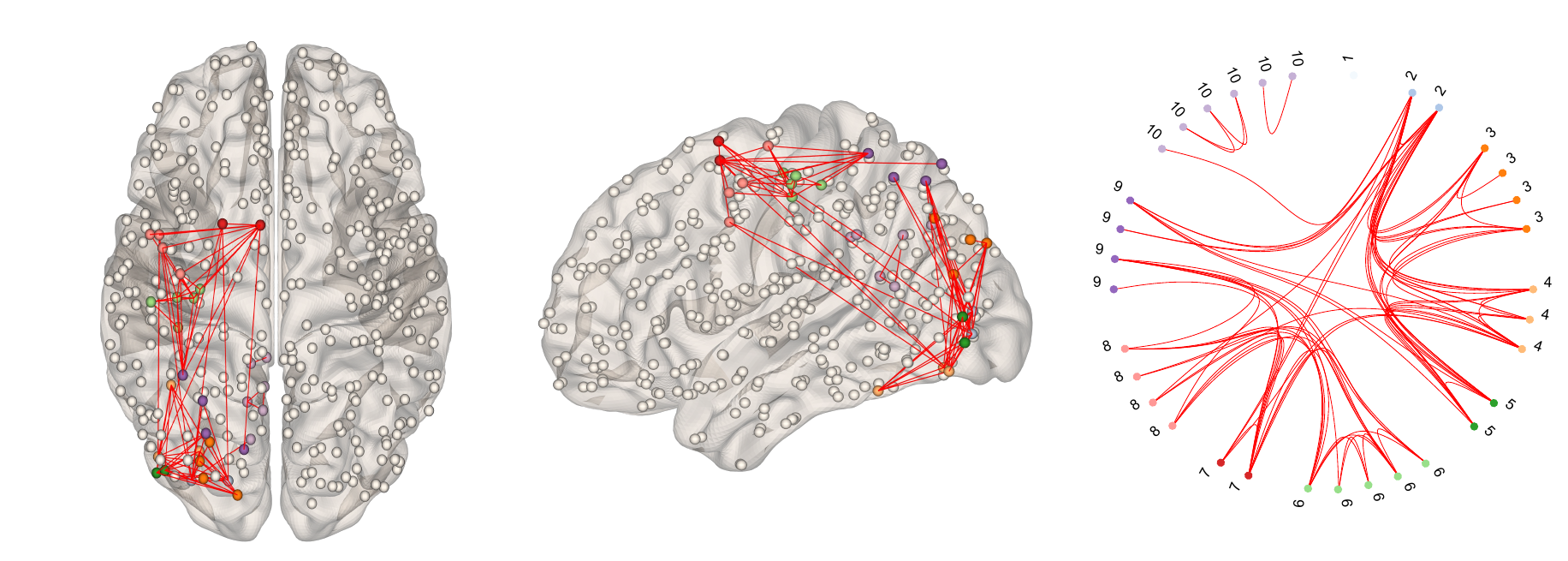}
\end{minipage}%
\hfill
\begin{minipage}[t]{0.20\textwidth}
    \centering
    \includegraphics[width=0.8\linewidth]{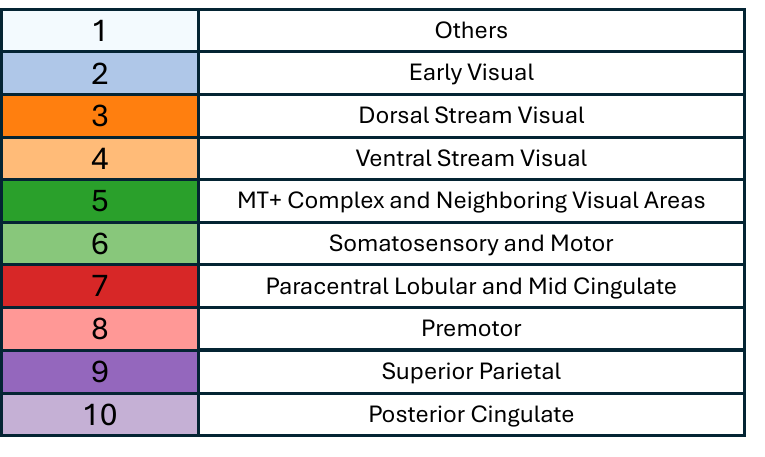}
\end{minipage}

\vspace{1ex}

\noindent
\textit{Axial and sagittal views of $\hat{\mathcal{B}}_1$ and the corresponding functional region network. Red edges indicate positive entries in $\hat{\mathcal{B}}_1$.}

\caption{Estimated $\hat{\mathcal{B}}_1$ from the external validation dataset.}
\label{fig:B_1}
\end{figure}

There is growing evidence that functional connectivity in the resting state exhibits remarkable self-organized fluctuations, which is relevant to understanding dementia progression  \citep{smith2013hcp}. To investigate this phenomenon, we analyze functional MRI (fMRI) data from the Human Connectome Project  \citep{vanessen2013hcp}, comprising 260 subjects who underwent both resting-state and task-based imaging sessions. During scanning, each subject completed five sequential stages, alternating between resting states and tongue-movement tasks. We extract blood-oxygen-level-dependent (BOLD) time series from 50 regions of interest (ROIs) across 30 time points. These time points are divided into five consecutive, non-overlapping time windows ($T = 5$), with the first and last windows corresponding to resting-state periods and the three middle windows to tongue-movement tasks. Within each time window, we use 6 time points to compute the sample covariance matrix of BOLD signals across the 50 ROIs. This results in a $50 \times 50$ covariance matrix $A_{it}$ representing the functional connectivity network for subject $i$ at time window $t$. We then fit the following matrix-response model:

\begin{equation}
  \boldsymbol{A}_{it}=\boldsymbol{\Theta} + \boldsymbol{\theta}_i + \mathcal{B}_1\text{Status}_{it} +
  \boldsymbol{\epsilon}_{it}, \\  
\end{equation}
where $\boldsymbol{\Theta}$ is the fixed intercept term, $\boldsymbol{\theta}_i$ is the random intercept,
$\text{Status}_{it}$ is the stage indicator variable ($\text{Status}_{it}=1$ for the action stage and $\text{Status}_{it}=0$
for the resting stage), and $\boldsymbol{\epsilon}_{it}$ is the error term. In our algorithm, the optimal values of tuning parameters $r$ and $s$ are selected using the eBIC criterion. We visualize the resulting $\hat{\boldsymbol{\Theta}}$ and $\hat{\mathcal{B}}_1$ in the same manner as described in Section \ref{sec:dti-imaging-data}, with the results displayed in Figures \ref{fig:theta} and \ref{fig:B_1}, respectively.

In Figure \ref{fig:theta}, the positive entries in Module 1 correspond to various visual modules, highlighting the significance of visual processing when subjects respond to visual cues for status changes. Module 2 encompasses the posterior cingulate cortex, a central hub in DMN. The positive values in this region underscore its critical function in sustaining elevated metabolic activity during resting states. Module 3, which incorporates the dorsal stream and superior parietal cortex, displays positive entries that support these areas' continuous engagement in spatial information processing, even when no specific task is required. Module 4 relates to somatosensory and motor-related brain regions, with positive values indicating maintained levels of motor readiness and sensory processing—reflecting the intrinsic functional state necessary for swift motor execution.

The results of $\hat{\mathcal{B}}_1$ presented in Figure \ref{fig:B_1} demonstrate a positive influence of
  tongue stimulation on the covariance between identified cortical regions. Positive entries corresponding to MT+ Complex
  and Neighboring Visual Areas, Dorsal Stream Visual, Early Visual, and Ventral Stream Visual indicate that the intrinsic
  connectivity within the visual processing network is preserved, potentially reflecting cross-modal interactions induced by
  tongue stimulation that operate alongside baseline visual system dynamics. Similarly, positive entries among somatosensory
  and motor regions, including Somatosensory and Motor and Premotor nodes, emphasize the sensorimotor system's capacity to
  maintain functional integration even when processing localized somatosensory input from the tongue. Furthermore, enhanced
  correlations among Posterior Cingulate (PC) nodes, a key component of DMN, suggest that
  intrinsic connectivity in this region remains intact despite external sensory stimulation. This observation is consistent
  with the PC's established role in integrating internal and external information. These findings lead us to conclude that
  tongue stimulation increases covariance between signals in the relevant ROIs, aligning with recent research on the effects
  of motor stimulation on brain connectivity. 



\section{Discussion}
\label{sec:discussion}
In this article, we have introduced a novel class of matrix-response generalized linear mixed models (MR-GLMMs)
specifically designed for analyzing longitudinal brain imaging data. Our approach addresses a critical gap in the
neuroimaging literature by providing a unified framework that can handle matrix-valued responses while accounting for
both between-subject heterogeneity through random effects and within-subject temporal correlations inherent in
longitudinal designs. The proposed MR-GLMM method extends traditional generalized linear mixed models to accommodate
matrix-valued responses through several key innovations. First, we model the conditional mean of the matrix response
using a linear combination of time-varying covariates with matrix-valued coefficients, incorporating both fixed
population-level effects and subject-specific random effects. To ensure computational tractability and statistical
interpretability, we impose a low-rank constraint on the population-level connectivity matrix through Burer-Monteiro
factorization, which is well-suited for brain network analysis where connectivity patterns often exhibit inherent
low-dimensional structure. Second, we enforce sparsity on the coefficient tensor using an $L_0$ constraint, reflecting
the biological reality that covariate effects are typically confined to specific subsets of brain connections rather
than affecting all edges uniformly. This sparsity assumption is crucial for interpretability in high-dimensional
neuroimaging applications. Third, we developed an efficient Monte Carlo Expectation-Maximization (MCEM) algorithm to
handle the computationally challenging optimization problem. The algorithm employs Metropolis-within-Gibbs sampling to
approximate the intractable integrals in the E-step, while the M-step utilizes proximal gradient descent with hard
thresholding to enforce the sparsity constraint on the coefficient tensor. The R code implementing our method is
available at \url{https://github.com/Zhentao20/MR_GLMM}.

While our proposed method represents an advancement, there are several limitations that suggest directions for future
research. Although our MCEM algorithm is more efficient than naive approaches, computational demands may become
prohibitive for very large-scale networks (e.g., whole-brain analyses with thousands of ROIs). Future work could explore
more efficient approximation schemes, such as variational inference methods or stochastic optimization algorithms that
could handle larger network dimensions while maintaining statistical rigor. Moreover, modern neuroscience increasingly
relies on multi-modal imaging approaches. Extending our framework to jointly model multiple types of connectivity data
(e.g., structural and functional networks) or to incorporate additional data types (e.g., genetic information,
behavioral measures) could provide more comprehensive insights into brain-behavior relationships.

\section{Acknowledgment}
Dr. Li's research is partially supported by NIH grant R01-AG073259. 

\section*{Appendix: Gradients Calculation}

This appendix details the gradient calculations for the objective function $F(\varthetab)$ of linear and logistic
models. For notational convenience, let
\[
\S = \boldsymbol{U}^\top \boldsymbol{U} - \boldsymbol{V}^\top \boldsymbol{V},
\text{ where }
S_{ks} = \sum_{a=1}^n (U_{ak}U_{as} - V_{ak}V_{as}).
\]



\paragraph{Linear model.} Define
\[
\eta^{(m)}_{it,jj'}
= (\boldsymbol{U}\boldsymbol{V}^\top)_{jj'} + \theta^{(m)}_{i,jj'}
+ \sum_{l=1}^p x_{itl} B_{jj'l},
\text{ and }
r^{(m)}_{it,jj'} = A_{it,jj'} - \eta^{(m)}_{it,jj'},
\]
where $\theta^{(m)}_{i,jj'}$ denotes the Monte Carlo samples as described in Section \ref{sec:monte-carlo-expect}.
Then, the elements of $\nabla F(\varthetab)$ are given by
\begin{align*}
\frac{\partial F}{\partial U_{js}}
&=
-\frac{1}{M}\sum_{m=1}^M \sum_{i=1}^N \sum_{t=1}^{T_i} \sum_{j'=1}^n
\frac{r^{(m)}_{it,jj'}}{\sigma_e^2} \, V_{j's}
+ 4\gamma \sum_{k=1}^r U_{jk} S_{ks}, \\
\frac{\partial F}{\partial V_{j's}}
&=
-\frac{1}{M}\sum_{m=1}^M \sum_{i=1}^N \sum_{t=1}^{T_i} \sum_{j=1}^n
\frac{r^{(m)}_{it,jj'}}{\sigma_e^2} \, U_{js}
- 4\gamma \sum_{k=1}^r V_{j'k} S_{ks}, \\
\frac{\partial F}{\partial B_{jj'l}}
&=
-\frac{1}{M}\sum_{m=1}^M \sum_{i=1}^N \sum_{t=1}^{T_i}
\frac{r^{(m)}_{it,jj'}}{\sigma_e^2} \, x_{itl}, \\
\end{align*}

\paragraph{Logistic model.} Define
\[
\eta^{(m)}_{it,jj'}
= (\boldsymbol{U}\boldsymbol{V}^\top)_{jj'} + \theta^{(m)}_{i,jj'}
+ \sum_{l=1}^p x_{itl} B_{jj'l},
\qquad
\pi^{(m)}_{it,jj'} =
\frac{\exp(\eta^{(m)}_{it,jj'})}{1+\exp(\eta^{(m)}_{it,jj'})},
\qquad
r^{(m)}_{it,jj'} = A_{it,jj'} - \pi^{(m)}_{it,jj'}.
\]
where $\theta^{(m)}_{i,jj'}$ denotes the Monte Carlo samples as described in Section \ref{sec:monte-carlo-expect}.
Then, the elements of $\nabla F(\varthetab)$ are given by
\begin{align*}
\frac{\partial F}{\partial U_{js}}
&=
-\frac{1}{M}\sum_{m=1}^M \sum_{i=1}^N \sum_{t=1}^{T_i} \sum_{j'=1}^n
r^{(m)}_{it,jj'} \, V_{j's}
+ 4\gamma \sum_{k=1}^r U_{jk} S_{ks}, \\
\frac{\partial F}{\partial V_{j's}}
&=
-\frac{1}{M}\sum_{m=1}^M \sum_{i=1}^N \sum_{t=1}^{T_i} \sum_{j=1}^n
r^{(m)}_{it,jj'} \, U_{js}
- 4\gamma \sum_{k=1}^r V_{j'k} S_{ks}, \\
\frac{\partial F}{\partial B_{jj'l}}
&=
-\frac{1}{M}\sum_{m=1}^M \sum_{i=1}^N \sum_{t=1}^{T_i}
r^{(m)}_{it,jj'} \, x_{itl}. \\
\end{align*}


\bibliographystyle{bibstyle}

\label{LastPage}
\end{document}